\documentclass[11pt,notoc]{JHEP3}

\usepackage{epsfig}
\usepackage{graphicx}
\usepackage{amssymb}
\usepackage{amsmath}
\usepackage{latexsym}

\DeclareMathOperator{\Tr}{Tr}

\title{(Reverse) Engineering Vacuum Alignment}
\author{Clifford Cheung and Jesse Thaler \\
Jefferson Physical Laboratory, Harvard University, Cambridge, MA
02138 \\ E-mail: \email{cwcheung@fas.harvard.edu},
\email{jthaler@jthaler.net} } \preprint{HUTP-06/A0015}

\abstract{In the presence of spontaneous symmetry breaking, the alignment of the
vacuum with respect to the gauge group is often controlled by
quadratically divergent operators in the low energy non-linear sigma
model. In principle the magnitudes and signs of these operators can
be changed by making different assumptions about the ultraviolet
physics, but in practice all known ways of regulating these theories
preserve the na\"{\i}ve vacuum alignment.  We show that by
``integrating in'' different sets of heavy spin-one fields, it is
possible to UV extend certain non-linear sigma models into two distinct UV insensitive theories.  These UV extensions have identical low energy degrees of freedom but different radiative potentials, making it possible to engineer two different vacuum alignments for the original non-linear sigma model.
Our construction employs ``non-square'' theory spaces which generically violate the common lore
that the preferred vacuum alignment preserves the
maximal gauge symmetry. By UV extending the $SO(9)/(SO(4)\times
SO(5))$ little Higgs model, we find a radiative potential that deviates from the na\"{\i}ve expectation but
does not stabilize the correct vacuum for proper
electroweak symmetry breaking.}

\begin{document}

\section{Motivation}
\label{sec:introduction}


Because a non-linear sigma model (NL$\Sigma$M) describes degrees of freedom whose properties
are defined entirely by spontaneous symmetry breaking, NL$\Sigma$Ms
are ideal for understanding the low energy physics of
QCD and technicolor, where strong dynamics obstructs calculability
and only the symmetries are known \emph{a priori}. Given the
variety of possible symmetry breaking patterns, NL$\Sigma$Ms have also
proven useful for building models with naturally light bosonic
states. From the earliest theories of composite Higgs bosons
\cite{CompositeHiggsScalars,SU(2)XSU(1)BreakingbyVacuumMisalignment}
to contemporary developments in little Higgs
\cite{EWSBfromDimensionalDeconstruction,MinimalMooseforaLH,LittlestHiggs,LHsfromanAntisymmetricCondensate,LHfromaSimpleGroup, SimpleModelofTwoLHs,LittlestHiggsModelwithCustodialSU(2)Symmetry,IntermediateHiggs},
$T$-parity \cite{Cheng:2003ju,Cheng:2004yc,TParityandtheLittlestHiggs}, holographic Higgs
\cite{HiggsasaHolographicPGB,MinimalCompositeHiggsModel}, and twin
Higgs theories
\cite{TwinHiggs:NaturalEWBfromMirrorSymmetry,NaturalLittleHierarchyfromaPartiallyGoldstoneTwinHiggs,TwinHiggsModelfromLeft-RightSymmetry},
NL$\Sigma$Ms have enriched our understanding of electroweak physics
beyond the standard model.

However, the problem of vacuum alignment can hinder the construction
of realistic models with the Higgs as a pseudo-Goldstone boson.
Consider a theory with a global symmetry $G$ spontaneously broken to
a subgroup $H\subset G$. If $G$ is an exact symmetry, then the
orientation of $H$ in $G$ is arbitrary, and the Goldstone bosons in
$G/H$ parameterize a space of equivalent vacua. If a subgroup
$F\subset G$ is weakly gauged, then gauge interactions specify a
second orientation in $G$ which lifts the vacuum degeneracy and
chooses a preferred alignment of $H$ relative to $F$. This vacuum
alignment is crucial to phenomenology because it sets the particle
content and charges of the low energy theory
\cite{AlignmentoftheVacuuminTheoriesofTC}. In both the $SU(6)/Sp(6)$
\cite{LHsfromanAntisymmetricCondensate} and $SO(9)/(SO(4)\times
SO(5))$ \cite{LittlestHiggsModelwithCustodialSU(2)Symmetry} little
Higgs theories, a na\"{\i}ve calculation indicates that the stable
vacuum does not allow for electroweak symmetry breaking.

Thus, it would be useful to have a method for engineering any
desired vacuum alignment for a $G/H$ NL$\Sigma$M with $F\subset G$
gauged. Of course, small tree level $G$-violating interactions can
force such an alignment by hand, but we are looking for a method to
fix the vacuum alignment from calculable one-loop corrections alone.
In particular, when a NL$\Sigma$M suffers from one-loop quadratic
divergences, one cannot trust the na\"{\i}ve stability of a given
vacuum alignment because it is controlled by UV sensitive operators
with incalculable coefficients.  Several remedies have been proposed
to render such operators calculable, all of which more or less
involve the addition of new spin-one resonances. For example, if the
theory is embedded in a higher dimensional spacetime, then what was
formerly a UV sensitive operator becomes a nonlocal Wilson loop in
the extra dimension. Since nonlocal operators cannot be
quadratically sensitive, KK loops end up softening divergences
\cite{Barbieri:2003pr}. Alternatively, one can invoke hidden local
symmetry
\cite{IstheRhoMesonaDynamicalGaugeBosonofHLS?,CompositeVectorMesonsfromQCDtoLH},
which introduces additional gauge bosons that cut off quadratically
divergent loop integrals.  As long as one assumes locality in theory
space \cite{VectorRealizationofChiralSymmetry}, then this
cancelation will occur \cite{Hirn:2004ze,Chivukula:2004kg}.



But in both cases, the calculable vacuum alignment is the same as
the vacuum alignment obtained from the na\"{\i}ve sign of
quadratically divergent operators
\cite{CompositeVectorMesonsfromQCDtoLH}. In QCD,  the
$\pi^{+}$ is heavier than the $\pi^{0}$, just as one would
``predict'' from the quadratically divergent photon loop, and regulating QCD with a $\rho$ meson in the vector limit \cite{VectorRealizationofChiralSymmetry} confirms that prediction \cite{Harada:2003xa}. There are examples like Casimir and thermal effects where the
na\"{\i}ve signs of quadratically divergent operators are just plain
incorrect. But in the context of vacuum alignment, we know of no
examples (until now) where different vacuum alignments can be
\emph{chosen} by making different assumptions about ultraviolet
physics.

In this paper, we present a novel method for UV extending low energy
NL$\Sigma$Ms  when the global symmetry breaking pattern is
\begin{equation}
U(N)\rightarrow U(M)\times U(N-M). \label{eq:masterbreak}
\end{equation}
There are two different ways of UV extending such
NL$\Sigma$Ms, and the two UV extensions are related by a
binary ``toggling'' operation which keeps the light
degrees of freedom fixed while reversing signs in the radiative
potential.  As we will see, the radiative potentials in these UV extensions are quadratically sensitive, but it is straightforward to further UV extend the NL$\Sigma$M into models that still exhibit toggling but are UV insensitive at leading order.  Because the original NL$\Sigma$M and its UV extensions have different gauge structures, the resulting radiative potentials are qualitatively different.  In contradiction to the common lore \cite{AlignmentoftheVacuuminTheoriesofTC}, the
stable vacuum alignment in non-square UV extensions does not always preserve
the maximal degree of gauge symmetry.

Toggling is only possible in ``non-square'' moose models. Until now,
research on mooses has centered on ``square'' moose models,
$\emph{i.e.}$ theory spaces in which every link field is a square
matrix whose left and right global symmetries act on the same number
of dimensions \cite{ToolkitforBuildersofCompositeModels}. One reason
for the popularity of square mooses is that they can be related to
extra-dimensional theories via deconstruction
\cite{(De)constructingDimensions}, and square link fields often have
natural UV completions as techni-fermion condensates.  In contrast,
because non-square mooses are crucial to the results of this paper,
our construction has no obvious extra-dimensional interpretation nor
straightforward UV completion using strong dynamics. Also, our
results depend crucially on theory space locality, which we assume
throughout this paper.

In Section \ref{sec:vendiagrams}, we discuss how two ultraviolet theories can have the same low energy degrees of
freedom but qualitatively different radiative potentials.  In
Section \ref{sec:twosite}, we UV extend NL$\Sigma$Ms based on
Eq.~(\ref{eq:masterbreak}) into two-site mooses that exhibit toggling.  Because these mooses are UV sensitive we cannot trust signs in the radiative potential, so
in Section \ref{sec:threesite} we present a simple example in which these theories are regulated into three-site mooses with finite one-loop potentials.  As expected, toggling between these three-site UV extensions does reverse the signs of pseudo-Goldstone masses.  In Section \ref{sec:tadpoles}, we study the
$SO(9)/(SO(4)\times SO(5))$ little Higgs model in hopes of
ameliorating its radiative instability, but discover that non-square mooses
introduce a pseudo-Goldstone tadpole not present in the original potential. We conclude with some
speculations about generalizing our construction to other
NL$\Sigma$Ms and UV completing non-square mooses.  Details of our calculations and further examples are given in the appendices.

\section{The Possibility of UV Extension}
\label{sec:vendiagrams}

\FIGURE[t]{\epsfig{file=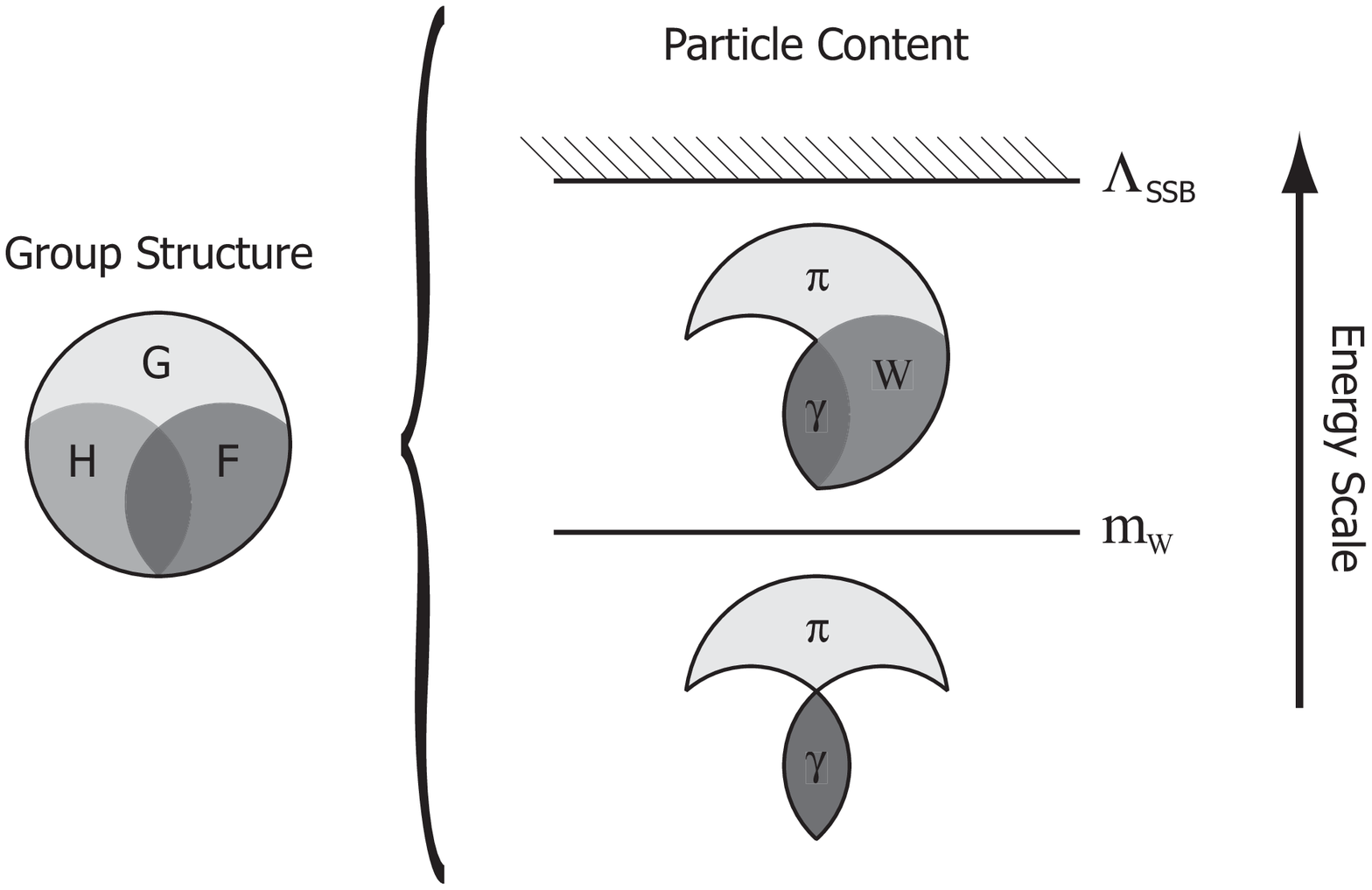,scale=0.5}
        \caption{Ven diagrams representing
the relative overlap of symmetry groups and particle multiplets.
$G$, $H$, and $F$ denote the global symmetry group, the unbroken
global symmetry subgroup, and the gauge group, respectively. Using
the language of technicolor, $\pi$, $\gamma$, and $W$ denote the
uneaten pseudo-Goldstones, the massless gauge bosons, and the
massive gauge bosons, respectively. At the scale $m_W$ it is
possible to produce $W$'s on-shell, and at $\Lambda_{\rm SSB}$ it is
necessary to introduce a UV completion to unitarize the NL$\Sigma$M.
} \label{VenDiagram1}}

Before focusing on a specific class of NL$\Sigma$Ms, we give a heuristic argument for how different UV assumptions can lead to different stable vacuum alignments. Consider a generic $G/H$ NL$\Sigma$M with a subgroup
$F\subset G$ weakly gauged. If we fix the alignment of $F$ relative
to $H$, then the theory consists of (pseudo-)Goldstone bosons in
$G/(F\cup H)$ (``$\pi$''), massless gauge bosons in $F\cap H$
(``$\gamma$''),
 and massive gauge bosons in $F/(F\cap H)$ (``$W$'') which acquire
 longitudinal modes via the Higgs mechanism.  We summarize the symmetries and degrees of freedom of this NL$\Sigma$M in Figure \ref{VenDiagram1}. Because the $W$ fields
are the heaviest degrees of freedom, we can integrate them out at
energies much lower than the symmetry breaking scale, leaving a low
energy theory comprised of just $\pi$ and $\gamma$. The alignment of
$F$ relative to $H$ is stable if there are no tachyonic
pseudo-Goldstone modes from radiative corrections.

\FIGURE[t]{\epsfig{file=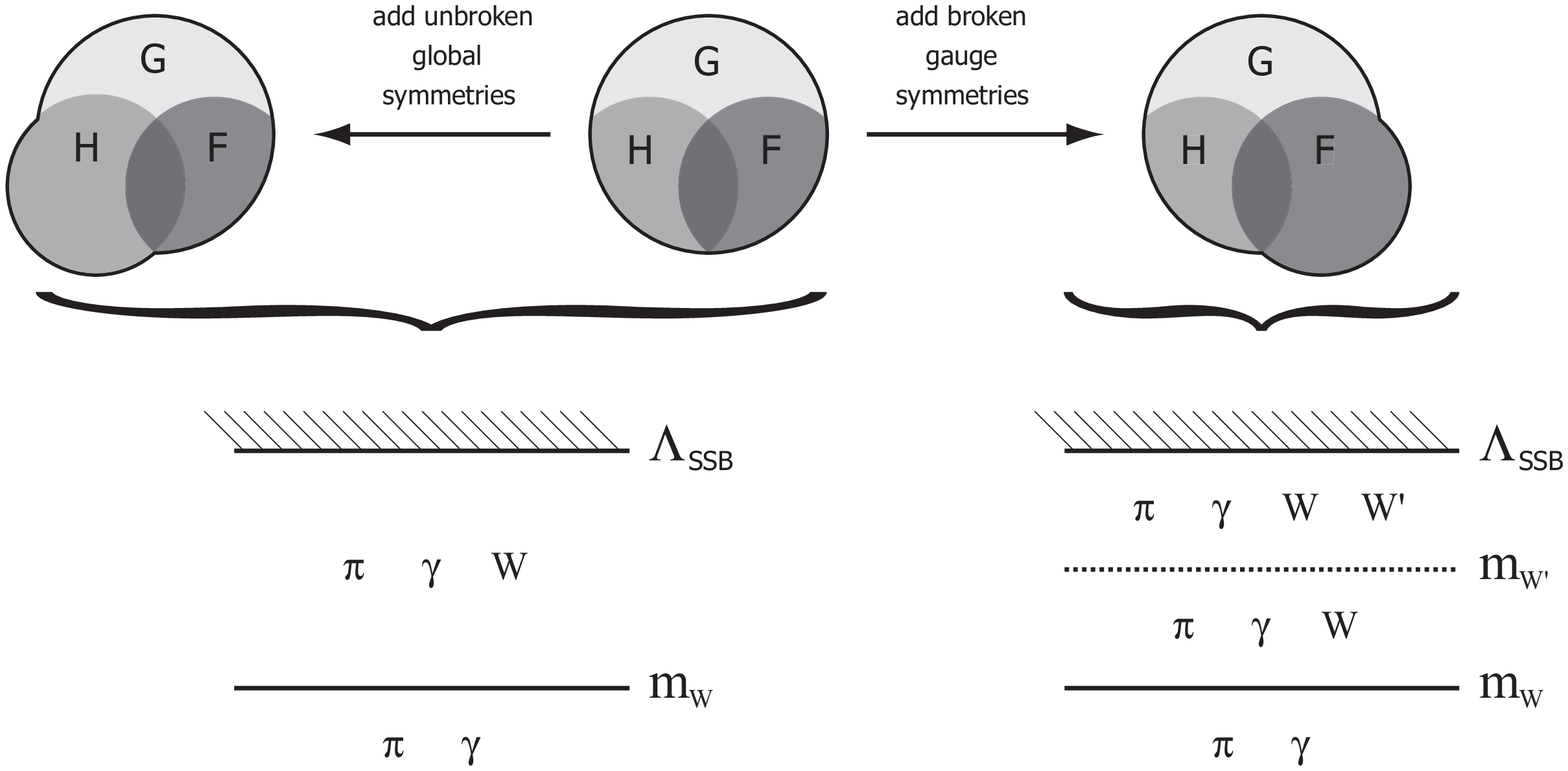,scale=0.5}
        \caption{Starting from a low energy
NL$\Sigma$M, one can generate new theories with identical
infrared physics either by 1) adding (or subtracting) unbroken
global symmetries, or 2) adding (or subtracting) broken gauge
symmetries. While the former leaves the particle content
unchanged all the way up to $\Lambda_{\rm SSB}$, the latter requires
the insertion of an intermediate scale, $m_{W'}$. While it is not
necessary that $m_{W'} > m_W$, we are imagining that $W'$'s gauge coupling is large compared to $W$'s.} \label{VenDiagram2}}

Under what conditions will two different NL$\Sigma$Ms share the
same low energy degrees of freedom? If we are only interested in
the dynamics of $\pi$ and $\gamma$ at tree level, then the
specific choices $G$, $H$, and $F$ are irrelevant as long as the
low energy gauge group $F\cap H$ and uneaten Goldstone fields
$G/(F\cup H)$ remain the same. This realization was used in
\cite{LittleTC} to show that the $SU(5)/SO(5)$ littlest Higgs
model could be UV completed into ordinary QCD with five flavors,
despite the fact that QCD has a chiral $SU(5)_{L}\times SU(5)_{R}$
global symmetry that is absent from the original model.

However, at one-loop level massive gauge bosons can affect the
radiative potential for pseudo-Goldstone bosons in the theory, and
depending on the specific choices of $G$, $H$, and $F$, there may or
may not be tachyonic modes. If all we do is enlarge $G$ by some new
global symmetry that is left unbroken by the vacuum then the
radiative potential will be the same because the gauge sector is
unchanged. If, on the other hand, we introduce a new global symmetry
that is fully gauged but maximally broken, then there is at least
the possibility that the radiative potentials will be different. We
call this a UV extension because the light degrees of freedom are
fixed but the structure of the radiative potential can change in the
presence of heavy gauge fields (see Figure \ref{VenDiagram2}).

\FIGURE[t]{\epsfig{file=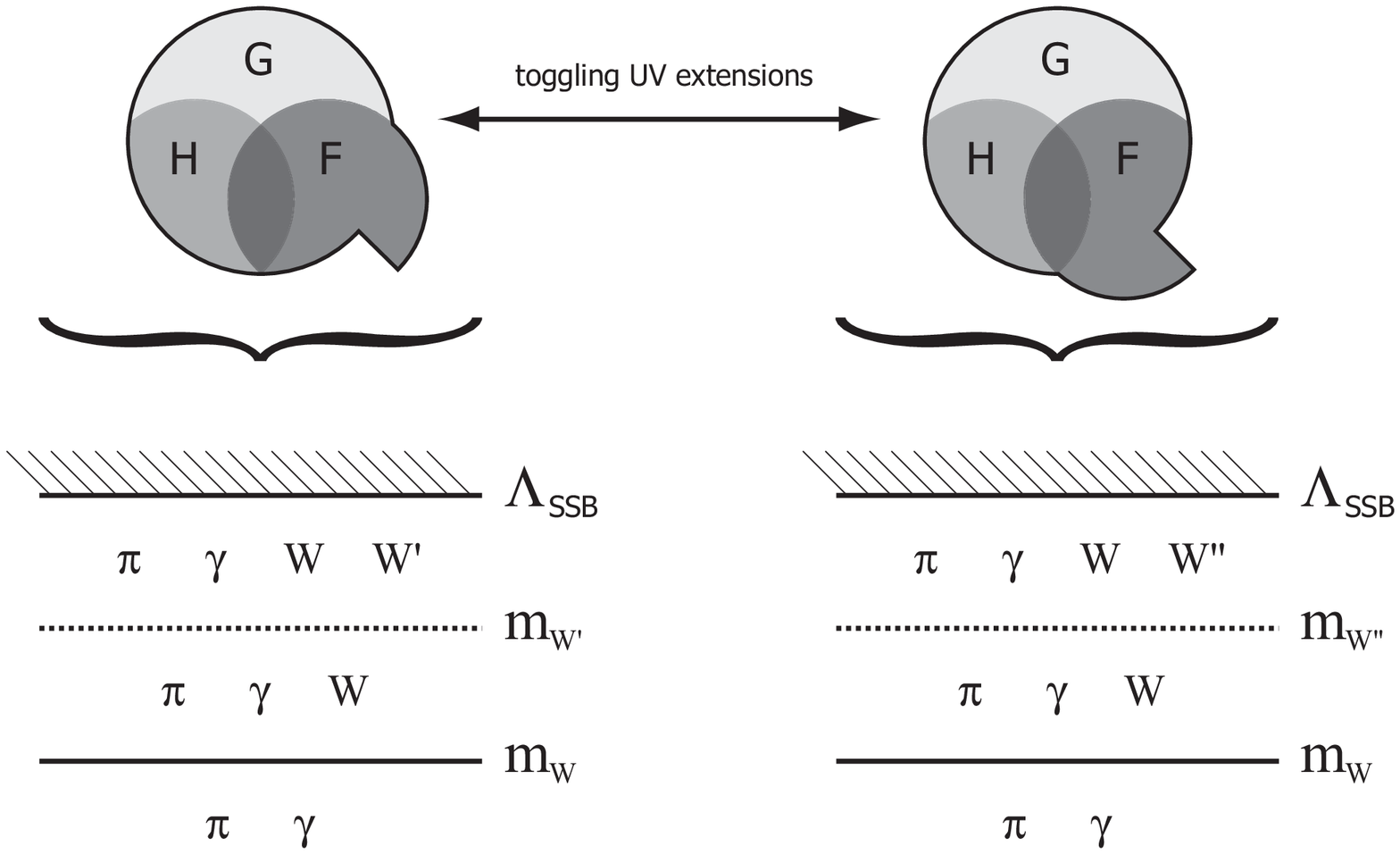,scale=0.5}
        \caption{Toggling is a binary
operation on theory space that interchanges two possible UV
extensions of the original NL$\Sigma$M. Depending on whether we
integrate in $W'$'s or $W''$'s above the scale $m_W$, we can
generate either ultraviolet theory, and the choice of UV extension
will affect the radiative stability of the vacuum.}
\label{VenDiagram3}}

Our claim is that NL$\Sigma$Ms based on
$U(N)\rightarrow U(M)\times U(N-M)$ admit two possible UV extensions with
different sets of heavy gauge bosons (see Figure \ref{VenDiagram3}).
As we describe in the following sections, the toggling operation that
interchanges these two UV extensions can be used to manipulate signs
in the radiative potential without changing the infrared particle
content.  In this way, different assumptions about heavy spin-one modes can lead to different stable vacuum alignments.

\section{Novel UV Extensions and Toggling}
\label{sec:twosite}


In this section we review little technicolor \cite{LittleTC} or hidden local symmetry \cite{IstheRhoMesonaDynamicalGaugeBosonofHLS?},
which can be used to UV extend any NL$\Sigma$M. Then we show
that for theories described by the symmetry breaking pattern
in Eq.~(\ref{eq:masterbreak}), there exist two novel UV extensions
using non-square mooses which are related by toggling.
While these UV extensions suffer from
quadratically sensitive operators, the na\"{\i}ve signs of these operators suggest
that toggling should flip signs in the radiative potential.  We will regulate this UV sensitivity in Section \ref{sec:threesite} and show that toggling still occurs in one-loop finite theories.

Using little technicolor we can UV extend a $G/H$ NL$\Sigma$M with $F\subset G$ gauged into a two-site
moose given by
\begin{equation}
\begin{tabular}{c}\includegraphics[clip,scale=0.4]{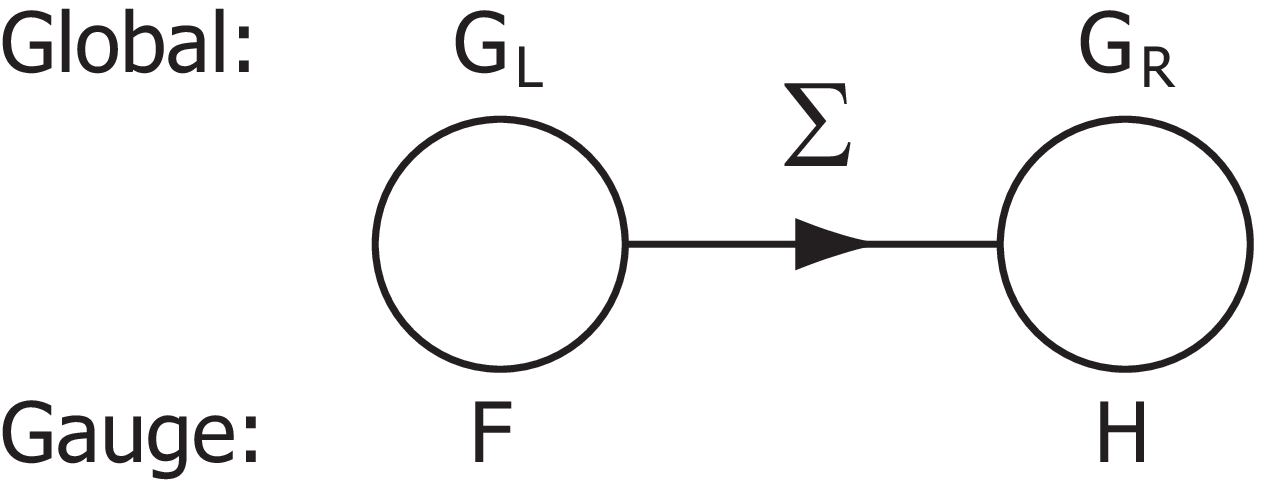}\end{tabular}
\qquad\qquad\begin{tabular}{c||c|c} & NL$\Sigma$M & Little
Technicolor \tabularnewline \hline \hline $\mathcal{G}$& $G$& $G_L\times
G_R$\tabularnewline \hline $\mathcal{H}$& $H$& $G_V$\tabularnewline \hline
$\mathcal{F}$& $F$& $F\times H$ \end{tabular}
\label{eq:newnotation}
\end{equation}
Because UV extending requires shuffling global and gauge symmetries,
we use a slightly different notation from Section
\ref{sec:vendiagrams}. Here $\mathcal{G}$, $\mathcal{H}$, and $\mathcal{F}$ denote the
complete global, unbroken, and gauge symmetries of an entire theory, while
$G$, $H$, and $F$ denote specific groups.

The symmetry structure of the above theory is compactly represented
by a moose diagram. Each site represents a symmetry group (global
and gauge symmetries denoted above and below the site, respectively)
while each link represents a matter field that transforms
bifundamentally under those symmetries.  For example, the link field
$\Sigma$ transforms as
\begin{eqnarray}
\Sigma \rightarrow L \Sigma R^{\dagger}, \qquad (L,R)\in& G_L\times
G_R.
\end{eqnarray}
As the result of spontaneous symmetry breaking, $\Sigma$ acquires a
vev equal to the identity matrix, leaving an unbroken symmetry
$G_V$. In the limit that $g_{H}$, the gauge coupling for $H$,
becomes large, the associated $H$ gauge fields become ultra-massive
and can be integrated out at low energies, yielding the original
NL$\Sigma$M.\footnote{While integrating out the $H$ gauge bosons
will generate a series of higher dimensional operators (such as four
fermion interactions and sigma field couplings), these terms will be
suppressed by coefficients given by na\"{\i}ve dimensional analysis
\cite{Manohar:1983md,Georgi:1986kr} where factors of $4 \pi$ are
replaced by $g_{H}$. See \cite{LittleTC} for details.}  The heavy
$H$ gauge bosons cut off one-loop quadratic divergences in the
radiative potential, and the qualitative structure of the regulated
radiative potential is the same as that of the original NL$\Sigma$M
\cite{CompositeVectorMesonsfromQCDtoLH}.  Moreover, it is clear from
the moose structure that little technicolor is the minimal
deconstruction of an extra dimension with bulk gauge symmetry $G$
bounded by two branes with reduced gauge symmetries $F$ and $H$, as
one would expect from the AdS/CFT correspondence
\cite{HiggsasaHolographicPGB}.

While the square moose little technicolor construction is
applicable to any NL$\Sigma$M, non-square moose UV
extensions can be implemented for NL$\Sigma$Ms of the
form
\begin{equation}
\mbox{NL$\Sigma$M: } \qquad
\begin{array}{ccc}
G & \; = \; &  U(N), \\
H & \; = \; & U(M)  \times U(N-M), \\
F & \; = \; & \emptyset. \label{eq:origNLSM}
\end{array}
\end{equation}
We have turned off gauge interactions $F$ for simplicity,
but as shown in Appendix \ref{sec:appendix1}, the following arguments still hold when they are included.  Until now, the only known two-site UV extension for the NL$\Sigma$M in
Eq.~(\ref{eq:origNLSM}) was
\begin{equation} \mbox{Little Technicolor: } \qquad
\begin{tabular}{c}\includegraphics[clip,scale=0.4]{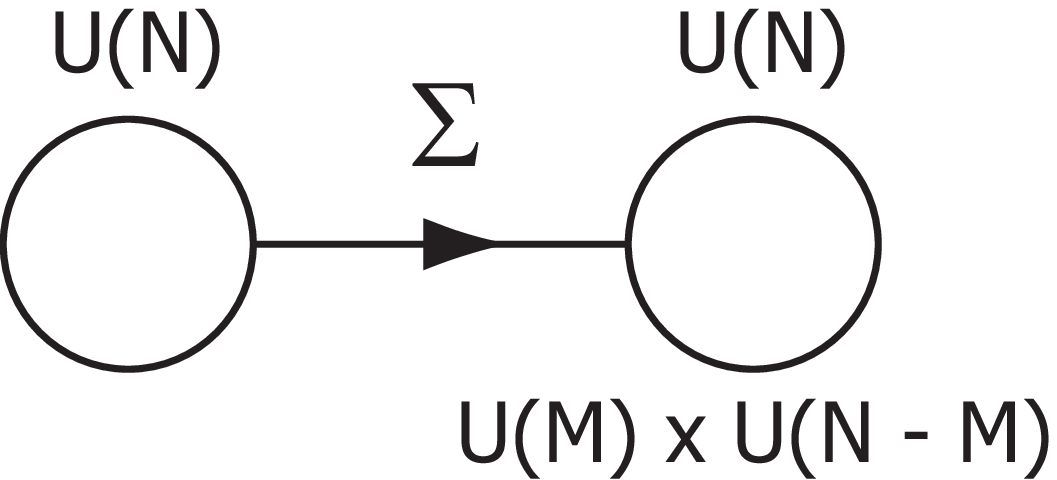}\end{tabular}
\qquad \langle\Sigma\rangle  =
\begin{array}{c} \left(\begin{array}{c}
\begin{array}{c}
\mathbf{1}_{N\times N}
\end{array}\end{array}\right)\end{array},
\label{eq:LTC}
\end{equation}
where $\mathbf{1}_{N\times N}$ denotes the $N\times N$ identity
matrix. As shown in Table \ref{table1}, the original NL$\Sigma$M and the little technicolor UV extension have the same low energy degrees of freedom.


\TABULAR[t]{c||c|c}{ & NL$\Sigma$M & Little Technicolor
\tabularnewline \hline \hline $\mathcal{G}$& $U(N)$& $U(N)\times
U(N)$\tabularnewline \hline $\mathcal{H}$& $U(M)\times U(N-M)$&
$U(N)$\tabularnewline \hline $\mathcal{F}$& $\emptyset$& $U(M)\times
U(N-M)$\tabularnewline \hline $\mathcal{G}/(\mathcal{F} \cup \mathcal{H})$&
$U(N)/\left(U(M)\times U(N-M)\right)$& $U(N)/\left(U(M)\times
U(N-M)\right)$\tabularnewline \hline $\mathcal{F}\cap \mathcal{H}$& $\emptyset$&
$\emptyset$\tabularnewline \hline \# of $\pi$'s& $2M(N-M)$&
$2M(N-M)$\tabularnewline \hline \# of $\gamma$'s& 0&
0\tabularnewline \hline \# of $W$'s& 0& $M^{2}+(N-M)^{2}$
\tabularnewline}{Symmetries of the original NL$\Sigma$M and
its little technicolor UV extension. Here we use the notation of
Eq.~(\ref{eq:newnotation}), where $\mathcal{G}$ denotes the global symmetry
group, $\mathcal{H}$ denotes the unbroken global symmetry subgroup, and
$\mathcal{F}$ denotes the gauge group. Note that while little technicolor
has more heavy gauge bosons compared to the NL$\Sigma$M, they have
the same light degrees of freedom. \label{table1}}

\TABULAR[t]{c||c|c}{& Theory A& Theory B \tabularnewline \hline \hline
$\mathcal{G}$& $U(N)\times U(M)$& $U(N)\times U(N-M)$\tabularnewline \hline
$\mathcal{H}$& $U(M)\times U(N-M)$& $U(M)\times U(N-M)$\tabularnewline
\hline $\mathcal{F}$& $U(M)$& $U(N-M)$\tabularnewline \hline
$\mathcal{G}/(\mathcal{F}\cup\mathcal{H})$& $U(N)/\left(U(M)\times U(N-M)\right)$&
$U(N)/\left(U(M)\times U(N-M)\right)$\tabularnewline \hline
$\mathcal{F}\cap\mathcal{H}$& $\emptyset$& $\emptyset$\tabularnewline \hline \#
of $\pi$'s& $2M(N-M)$& $2M(N-M)$\tabularnewline \hline \# of
$\gamma$'s& 0& 0\tabularnewline \hline \# of $W$'s& $M^{2}$&
$(N-M)^{2}$\tabularnewline}{Symmetries of Theories A and B.
Both of theories have the same infrared particle content as the
original NL$\Sigma$M. \label{table2}}

Non-square mooses permit two novel UV extensions of the original
NL$\Sigma$M, denoted by Theory A and Theory B.  Just like little
technicolor, these UV extensions have the same low energy Lagrangian
as the original NL$\Sigma$M after integrating out the extra massive
gauge bosons.   The moose and vacuum structure of Theory A
is
\begin{equation} \mbox{Theory A: } \qquad
\begin{tabular}{c}\includegraphics[clip,scale=0.4]{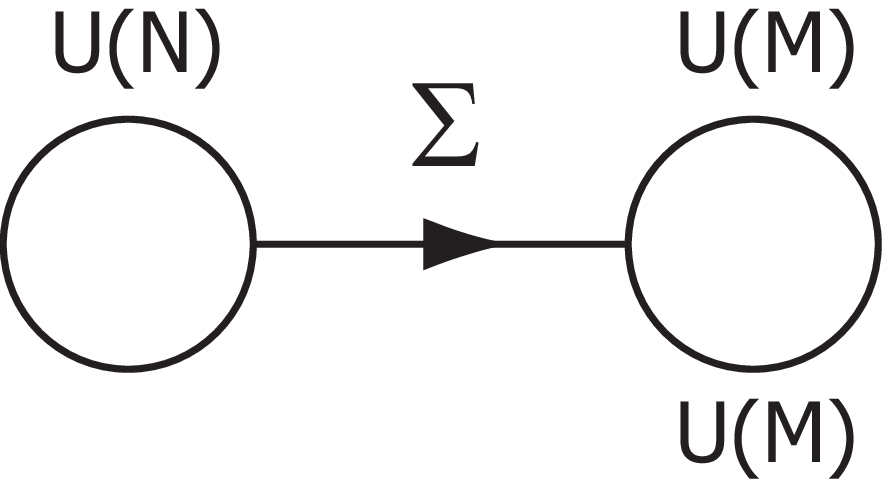}\end{tabular}
\qquad\langle\Sigma\rangle  =
\begin{array}{c} \left(\begin{array}{c}
\begin{array}{c}
\mathbf{1}_{M\times M}\\
\mathbf{0}_{(N-M)\times M}\end{array}\end{array}\right)\end{array},
\label{eq:TI}
\end{equation}
where $\mathbf{0}_{(N-M)\times M}$ denotes the $(N-M)\times M$ zero
matrix.  We can generate Theory B by toggling Theory A.  For any non-square
moose, toggling has a two-fold effect: (a) it replaces a fully
gauged symmetry at a site with a ``conjugate'' fully gauged
symmetry, and (b) it replaces the vev of the neighboring link field
with the ``conjugate'' vev.  In particular, toggling sends Theory A
to Theory B by replacing $U(M)$ with $U(N-M)$, giving
\begin{equation} \mbox{Theory B: } \qquad
\begin{tabular}{c}\includegraphics[clip,scale=0.4]{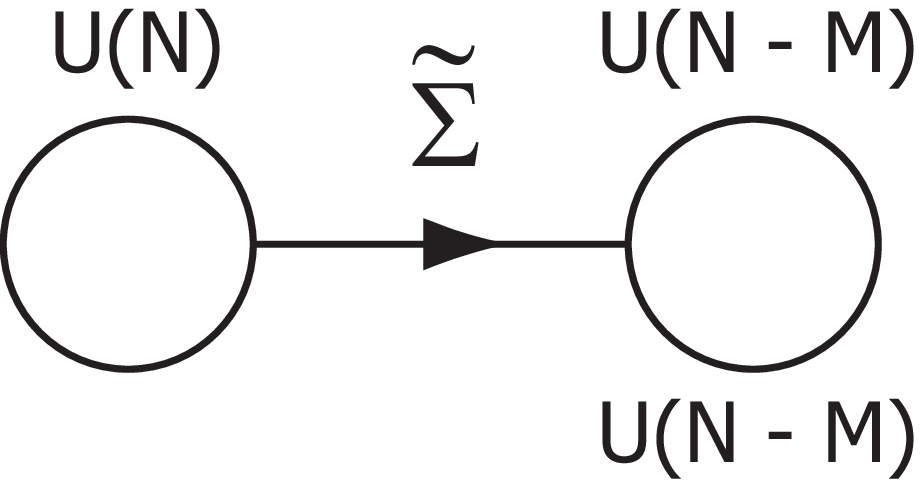}\end{tabular}
\qquad \langle\tilde{\Sigma}\rangle =
\begin{array}{c} \left(\begin{array}{c}
\begin{array}{c}
\mathbf{0}_{M\times(N-M)}\\
\mathbf{1}_{(N-M)\times(N-M)}\end{array}\end{array}\right)\end{array}.
\label{eq:TII}
\end{equation}
As shown in Table \ref{table2}, Theories A and B also
have same light degrees of freedom as the original NL$\Sigma$M, and this low energy equivalence is calculated explicitly in Appendix \ref{sec:appendix1}.

From the form of the vevs in Eqs.~(\ref{eq:TI}) and
(\ref{eq:TII}), note that
\begin{equation}
\langle\tilde{\Sigma}\rangle\langle\tilde{\Sigma}\rangle^{\dagger}=1-\langle\Sigma\rangle\langle\Sigma\rangle^{\dagger}.
\end{equation}
As shown in Appendix \ref{sec:appendix1}, in
unitary gauge $\Sigma$ and $\tilde{\Sigma}$ are written in
terms of the \emph{same} Goldstone matrix $\pi$, so
\begin{equation}
\Sigma = e^{i\pi/f}\langle\Sigma\rangle, \qquad \tilde{\Sigma} =
e^{i\pi/f} \langle\tilde{\Sigma}\rangle,
\end{equation}
where $f$ is the link field decay constant.  Therefore, as far as the radiative potentials are concerned, toggling sends
\begin{equation}
\label{eq:mastertoggle}
\Sigma \Sigma^\dagger \rightarrow
\tilde{\Sigma}\tilde{\Sigma}^\dagger = 1 - \Sigma \Sigma^\dagger,
\end{equation}
where the minus sign in the last term will be crucial to
manipulating radiative potentials in the following
sections.



In the presence of $F$ gauge interactions, the light degrees of
freedom in little technicolor and Theories A and B are exactly same
as those in the original NL$\Sigma$M with $F$ gauged. While these theories have
the same infrared particle content, they do not
have the same radiative potentials since they have
different gauge structures.  In terms of the
gauge boson mass matrix
\begin{eqnarray}
(M^2)^{ab} & = & \frac{\partial^2 \mathcal{L}}{\partial A_\mu^a
\partial A^{\mu b}},
\label{eq:massmatrix}
\end{eqnarray} the one-loop Coleman-Weinberg radiative potential
\cite{RadiativeCorrectionsastheOriginofSSB} is
\begin{eqnarray}
V_{{\rm CW}} & = & \frac{3\Lambda^{2}}{32\pi^{2}}{\rm
Tr}(M^{2})+\frac{3}{64\pi^{2}}\Tr\left(M^4 \log
\frac{M^2}{\Lambda^{2}} \right).
\label{eq:CW}
\end{eqnarray} Using this formula, both Theory A
and Theory B have UV sensitive operators of the form
\begin{equation}
\label{eq:quaddivop}
\Lambda^{2}\Tr(\Sigma\Sigma^{\dagger}C_F),
 \end{equation}
where $C_F = T_{F}^{a}T_{F}^{a}$ is the quadratic Casimir of $F$ and $\Sigma\Sigma^{\dagger}$
is \emph{not} the identity because $\Sigma$ is not a square unitary
matrix. Thus, strictly speaking the vacuum alignments in Theories A and B are unknown, as summarized in Figure
\ref{BigPicture}.  Because Eq.~(\ref{eq:quaddivop}) could never arise in any square moose radiative potential, non-square theories have novel potentials compared to standard UV extensions.

\FIGURE[t]{\epsfig{file=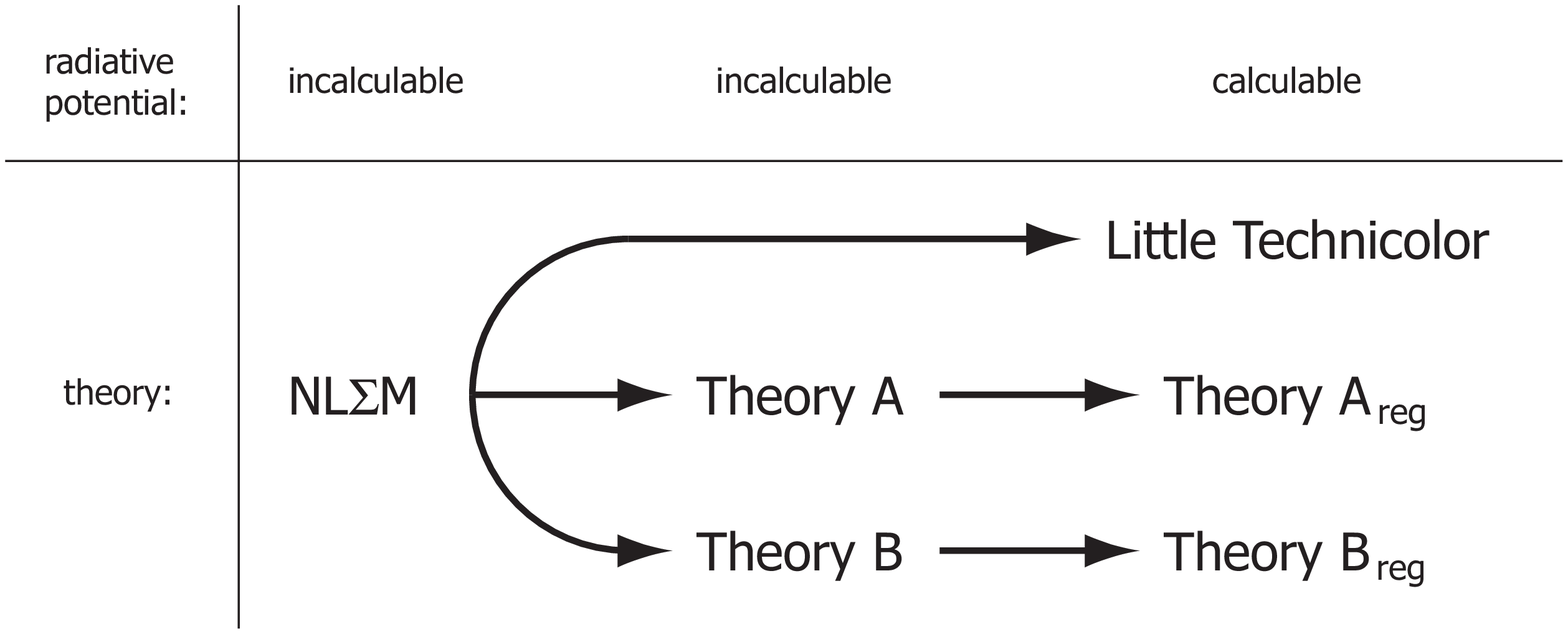,scale=0.5}
        \caption{A summary of UV extensions of the original NL$\Sigma$M,
        specifying moose structure and radiative stability of the
na\"{\i}ve vacuum alignment.  Because Theories A and B receive
quadratically divergent radiative corrections, the signs of UV
sensitive operators are incalculable. Nonetheless, these divergences
can be rendered finite (and thus calculable) by UV extending into
Theories A$_{\rm reg}$ and B$_{\rm reg}$ as discussed in Section
\ref{sec:threesite}.  In the little technicolor UV extension, the
radiative potential is still calculable, albeit logarithmically
divergent.} \label{BigPicture}}

Although the coefficient in front of Eq.~(\ref{eq:quaddivop}) should
not be taken seriously, it is interesting to note that
Eq.~(\ref{eq:mastertoggle}) sends
\begin{eqnarray}
 & \Lambda^{2}\Tr(\Sigma\Sigma^{\dagger}C_F) \rightarrow
 \Lambda^{2}\Tr(\tilde{\Sigma}\tilde{\Sigma}^{\dagger}C_F) =
 -\Lambda^{2}\Tr(\Sigma\Sigma^{\dagger}C_F)+\rm{constant}. \label{eq:toggcon} \end{eqnarray}
Our na\"{\i}ve analysis indicates that modulo a constant, toggling
flips the sign of this operator in the radiative potential.
Physically, this difference occurs because the $U(M)$ and $U(N-M)$
gauge fields in Theories A and B form distinct subsets whose union
comprises the full $U(M)\times U(N-M)$ of little technicolor. Since
little technicolor has a constant one-loop quadratic divergence, the
$\Sigma$ dependent contributions from the $U(M)$ and $U(N-M)$ gauge
multiplets must cancel, and thus Theories A and B have nontrivial
quadratic divergences that differ only by a sign.\footnote{It is
important to note that this property only holds for quadratically
sensitive operators, and is a consequence of the fact that the
Casimir of a product group $H = H_1 \times H_2$ is the sum of the
Casimirs of $H_1$ and $H_2$.  For the calculable contributions to
the radiative potential, Eq.~(\ref{eq:quadsum}) no longer holds.}
Roughly speaking, at the level of one-loop UV sensitive operators
\begin{equation}
\label{eq:quadsum}
V_{\text{Theory A}} + V_{\text{Theory B}} =
V_{\text{Little Technicolor}},
\end{equation}
where the right hand side is precisely the constant in
Eq.~(\ref{eq:toggcon}).

One can still ask whether this result holds if we properly regulate
these UV sensitive theories. In other words, if we UV extend
Theories A and B into healthy, quadratic divergence-free theories,
will toggling flip the signs of radiatively generated operators? The
answer is yes, and we will consider a concrete example in the next
section.

\section{A Simple Example with UV Insensitive Toggling}
\label{sec:threesite}

Take the
NL$\Sigma$M from the previous section with $N=3$, $M=2$, and a gauged $U(2)$ subgroup:
\begin{equation}
\begin{array}{ccc}
G & \; = \; &  U(3), \\
H & \; = \; & U(2)  \times U(1), \\
F & \; = \; & U(2)
\end{array}
\label{eq:simplest}
\end{equation}
This NL$\Sigma$M has roughly the same low energy degrees of freedom as the simple group little Higgs \cite{LHfromaSimpleGroup}.
In Section \ref{sec:twosite}, we learned that this theory and its UV
extensions, Theories A and B, exhibit quadratically divergent
radiative corrections. In this section, we will regulate these UV
sensitive operators by introducing even more heavy spin-one
resonances. The resulting UV extensions, denoted by Theory A$_{\rm
reg}$ and Theory B$_{\rm reg}$, will have the same low energy degrees of
freedom as the original NL$\Sigma$M, but qualitatively different radiative structures. Moreover, in accordance with
the na\"{\i}ve prediction made in the previous section, these
theories will have opposite radiative stability, as summarized in Figure
\ref{ThreeSiteEquiv}.

\FIGURE[t]{\epsfig{file=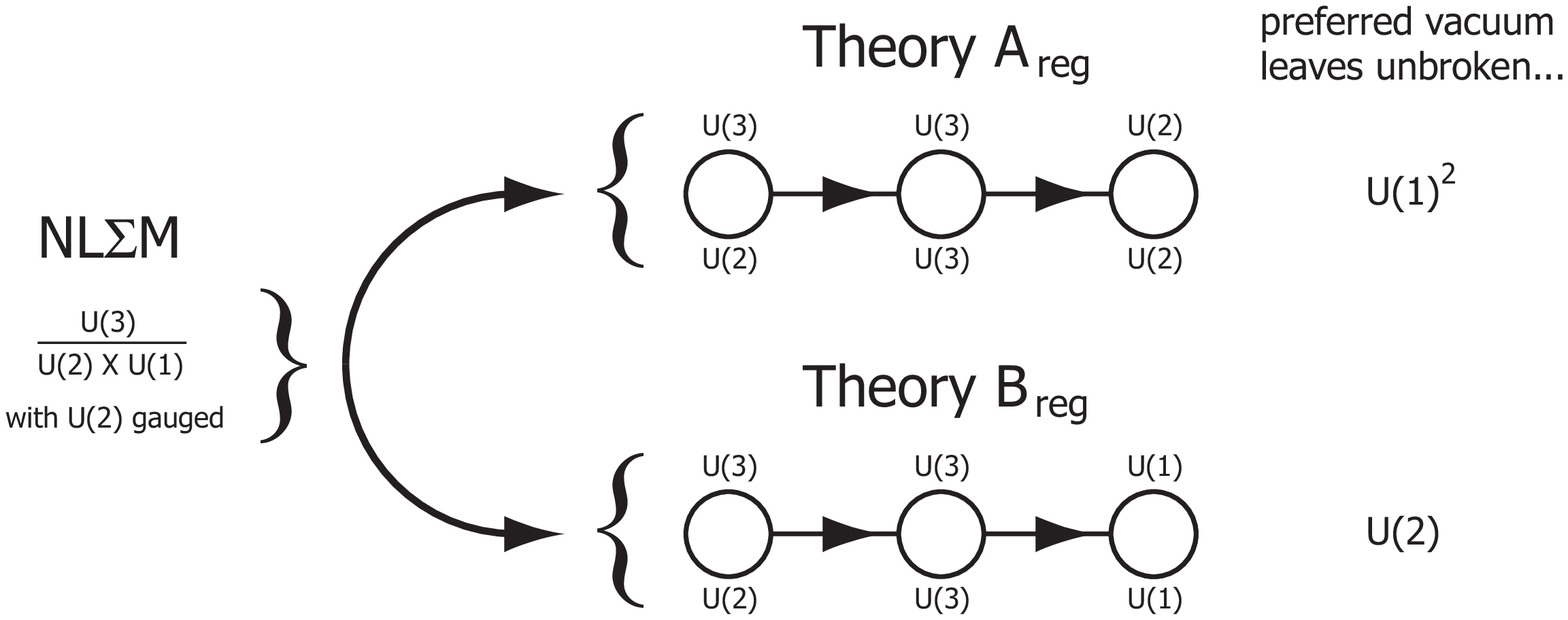,scale=0.5}
        \caption{A summary of the regularized UV extensions of the NL$\Sigma$M in Eq.~(\ref{eq:simplest}), specifying moose structure and the gauge group preserved by the
stable vacuum alignment. Interestingly, the stable vacuum for Theory
A$_{\rm reg}$ is not that which preserves the maximal amount of
gauge symmetry.} \label{ThreeSiteEquiv}}

By inserting a fully gauged $U(3)$ site in the center of the moose in Eq.~(\ref{eq:TI}), it is
possible to regularize the quadratic divergences in Theory A:
\begin{equation} \mbox{Theory A$_{\rm reg}$: } \qquad
\begin{tabular}{c}\includegraphics[clip,scale=0.4]{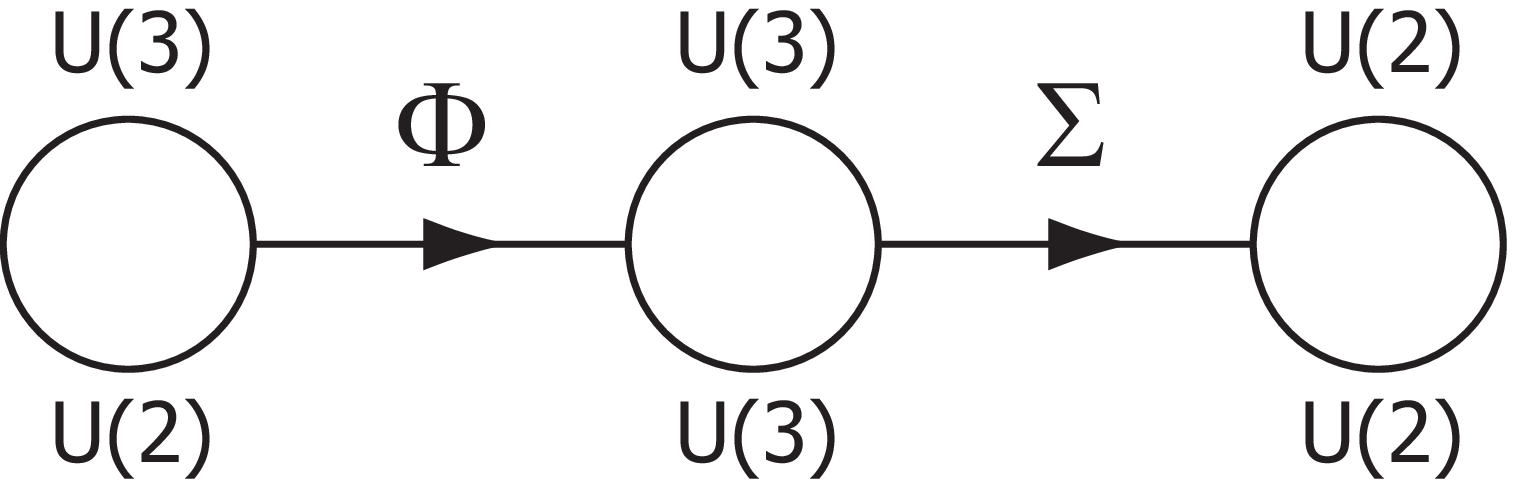}\end{tabular}
\label{eq:Areg}
\end{equation}
\begin{eqnarray}
\langle\Phi\rangle=\left(\begin{array}{ccc}
1 & 0 & 0\\
0 & 1 & 0\\
0 & 0 & 1\end{array}\right), & \qquad &
\langle\Sigma\rangle=\left(\begin{array}{cc}
1 & 0\\
0 & 1\\
0 & 0\end{array}\right), \label{eq:unstablevev}\end{eqnarray}
where $\Phi$ is an ordinary square link field.  Likewise, Theory B in Eq.~(\ref{eq:TII}) can be UV extended into
\begin{equation} \mbox{Theory B$_{\rm reg}$: } \qquad
\begin{tabular}{c}\includegraphics[clip,scale=0.4]{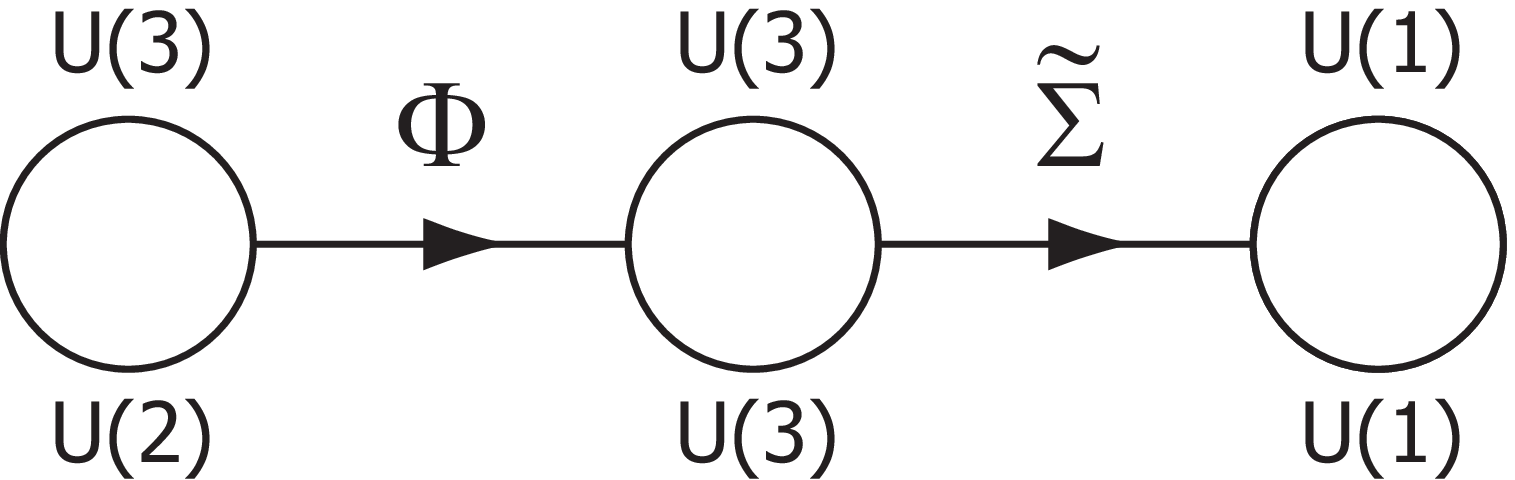}\end{tabular}
\label{eq:Breg}
\end{equation}
\begin{eqnarray}
\langle\Phi\rangle=\left(\begin{array}{ccc}
1 & 0 & 0\\
0 & 1 & 0\\
0 & 0 & 1\end{array}\right), & \qquad &
\langle\tilde{\Sigma}\rangle=\left(\begin{array}{c}
0\\
0\\
1\end{array}\right).
\end{eqnarray}
Because we can go to a gauge where all the Goldstones in $\Phi$ are eaten by the $U(3)$ gauge group, it is obvious that Theories A$_{\rm reg}$ and B$_{\rm reg}$ have the same light degrees of freedom as Theories A and B and are therefore valid UV extensions of the original NL$\Sigma$M in Eq.~(\ref{eq:simplest}).

While the number of massive gauge bosons are different in these five
theories, the light degrees of freedom are the same after
integrating out heavy modes:  they all contain a scalar doublet
$h$ charged under an unbroken $U(2)$ gauge symmetry.  In unitary
gauge, the link fields in Theories A$_{\rm reg}$ and B$_{\rm reg}$
are
\begin{eqnarray}
\Phi = e^{i\pi/f_{\rm eff}}\langle\Phi\rangle, \qquad \Sigma =
e^{i\pi/f_{\rm eff}}\langle\Sigma\rangle, \qquad \tilde{\Sigma} =
e^{i\pi/f_{\rm eff}}\langle\tilde{\Sigma}\rangle,
\label{eq:goldstonepar}
\end{eqnarray}
where the Goldstone matrix is
\begin{eqnarray} \pi & = & \frac{1}{\sqrt{2}}
\left(\begin{array}{cc}
\begin{array}{cc}
0 & 0\\
0 & 0\end{array} & h\\
\begin{array}{c}
h^{\dagger}\end{array} & 0\end{array}\right),\end{eqnarray} and the
effective pion decay constant is defined to properly normalize the
Goldstone kinetic terms:
\begin{equation}
f_{\rm eff} = \sqrt{f_\Phi^2 + \frac{f_\Sigma^2}{2}}.
\label{eq:feff}
\end{equation}

It is now straightforward to calculate the radiative potentials for
Theories A$_{\rm reg}$ and B$_{\rm reg}$ to see the effect of
toggling. Since ${\rm Tr}(M^2)$ and ${\rm Tr}(M^4)$ are constants,
both theories are completely free from quadratic and logarithmic
divergences at one-loop.

The cancelation of divergences can be better understood in the
language of spurions.  In the absence of gauge couplings, Theory
A$_{\rm reg}$ has three non-linearly realized $U(3)$ symmetries
which protect the Goldstones from a mass: two acting on $\Phi$ from
the left and right, and one acting on $\Sigma$ from the left (the
$U(2)$ acting on $\Sigma$ from the right does not forbid mass
terms). Denoting the gauge couplings on the left, middle and right
gauge sites by $g_F$, $g_G$, and $g_H$, we see that $g_F$ breaks the
$U(3)$ acting on $\Phi$ from the left and $g_G$ breaks the other two
$U(3)$'s to the diagonal. Consequently, masses for the Goldstones
must be generated at order $g_F g_G^2$.\footnote{In
Eqs.~(\ref{eq:higgs1}) and (\ref{eq:higgs2}) below, the extra factor
of $g_G^2$ comes from the fact that the $\alpha_i$ are proportional
to $g_G^2$. The fact that $g_G$ controls the breaking of two
different symmetries is reminiscent of collective breaking in the
simple group little Higgs \cite{LHfromaSimpleGroup}.} Note that this
differs from little technicolor, which has only two $U(N)$ chiral
symmetries protecting the Goldstone masses, and thus has
logarithmically divergent radiative corrections.

To compute the finite Coleman-Weinberg radiative potential it is
convenient to sum over vacuum bubbles in a background of
pseudo-Goldstone bosons
\begin{equation}
V_{\rm CW} = \frac{3}{2}{\rm Tr}\int_0^\infty \frac{p^3 dp}{8
\pi^2}\log\left(1+\frac{M^2}{p^2}\right),\label{eq:CW2}
\end{equation}
where $M^2$ is the gauge boson mass matrix and we have analytically
continued into Euclidean space. Using the matrix identity ${\rm
Tr}\log X = \log {\rm Det} X$ and expanding to quadratic order in
the Goldstones, we find that to leading order in $g_F^2$, the mass
of the Higgs mode in Theory A$_{\rm reg}$ is
\begin{eqnarray}
\label{eq:higgs1}
m_h^2 &=& -\frac{3}{32\pi^2}\frac{f_\Phi^2 f_\Sigma^2 g_F^2 g_G^2}{f_{\rm eff}^2}  \log \left( m_1^{\alpha_1}m_2^{\alpha_2}m_3^{\alpha_3} \right) +
\mathcal{O}(g_F^4).
\end{eqnarray}
As detailed in Appendix \ref{sec:appendix2}, $m_1 > m_2 > m_3$ are
physical gauge boson masses and $\alpha_i$ are coefficients that
satisfy $\sum_i \alpha_i  =0$.  For Theory B$_{\rm reg}$ we find
\begin{eqnarray}
\label{eq:higgs2}
\tilde{m}_h^2 &=& \frac{3}{32\pi^2}
\frac{f_\Phi^4 g_F^2 g_G^2}{f_{\rm eff}^2}\log
\left(\frac{\tilde{m}_1}{\tilde{m}_2} \right) + \mathcal{O}(g_F^4),
\end{eqnarray}
where $\tilde{m}_1 >\tilde{m}_2$.  Note that $\tilde{m}_h^2$ is
manifestly positive.  Moreover, if we demand that the gauge
couplings and pion decay constants are consistent with Theory
A$_{\rm reg}$ being a UV extension of Theory A (\emph{i.e.}\ such
that new massive gauge bosons are heavier than pre-existing ones),
then $m_h^2$ is negative.  For example, given $f_1=f_2=f$
and $g_G=g_H=g$,
 \begin{eqnarray} m_h^2 &=& \frac{f^2
g_F^2 g^2}{16\pi^2}\left(\log\left(\frac{3}{2}\right) +
\frac{\sqrt{5}}{20} \log\left(
\frac{47-21\sqrt{5}}{2}\right)\right)+\mathcal{O}(g_F^4)<0 ,\\
\tilde{m}_h^2 &=& \frac{f^2 g_F^2
g^2}{16\pi^2}\log\left(\frac{3}{2}\right) + \mathcal{O}(g_F^4)>0.
\end{eqnarray}
Thus $m_h^2<0<\tilde{m}_h^2$, which is consistent with the
na\"{\i}ve signs given by Eq.~(\ref{eq:quaddivop}) for Theories A
and B.  Our prediction from Section \ref{sec:twosite} holds for
Theories A$_{\rm reg}$ and B$_{\rm reg}$ and toggling does indeed
flip signs in the radiative potential.

Alternatively, imagine starting with a low energy theory comprised
of an $h$ doublet and a $U(2)$ gauge symmetry. We can UV extend this
theory into Theories A$_{\rm reg}$ and B$_{\rm reg}$ by integrating
in heavy spin-one modes. Since these theories have
different radiative potentials, we have the freedom to UV extend
into whichever theory has the desired radiative stability. If we want a tachyonic
doublet for ``electroweak'' symmetry breaking, then we
would choose Theory A$_{\rm reg}$, which radiatively generates a
negative mass squared for $h$. If we instead want a stable doublet,
we would choose Theory B$_{\rm reg}$. In this way, it is
possible to (reverse) engineer vacuum alignment by making different
assumptions about the ultraviolet physics.

We have established that the vacuum alignment we chose for Theory
A$_{\rm reg}$ is unstable, but what is the stable vacuum alignment?
As it turns out, it is
\begin{eqnarray}
\langle\Phi\rangle=\left(\begin{array}{ccc}
1 & 0 & 0\\
0 & 1 & 0\\
0 & 0 & 1\end{array}\right), &  \qquad &
\langle\Sigma\rangle=\left(\begin{array}{cc}
0 & 0\\
1 & 0\\
0 & 1\end{array}\right).\end{eqnarray} This stable vacuum
alignment yields an unbroken $U(1)^2$ gauge symmetry, which has
fewer generators than the unbroken $U(2)$ that results from the
unstable vacuum alignment defined in Eq.~(\ref{eq:unstablevev}).
Thus, despite the common lore, the vacuum does not necessarily align
to preserve the maximal unbroken gauge symmetry. As we show in
Appendix \ref{sec:appendix3}, non-square moose models violate this
expectation quite generically.


\section{UV Extending the Littlest Higgs with Custodial Symmetry}
\label{sec:tadpoles}

We have seen that for certain UV sensitive NL$\Sigma$Ms, one can
generate two different UV extensions with finite one-loop potentials.  Moreover,
toggling between these theories can flip signs in the radiative
potential without changing the low energy degrees of freedom. Can this construction be used to
stabilize vacuum alignments in phenomenologically interesting
theories? Since the arguments made in Section \ref{sec:twosite} and
Section \ref{sec:threesite}  hold equally for orthogonal groups as
well as as unitary groups, we can apply our results to the
$SO(9)/(SO(4)\times SO(5))$ little Higgs model, described in
\cite{LittlestHiggsModelwithCustodialSU(2)Symmetry}.

In this theory,
the vacuum spontaneously breaks a global $SO(9)$ to $SO(4)\times
SO(5)$ with $SU(2)^3 \times U(1)\subset SO(9)$ gauged, leaving
massless $SU(2)_L\times U(1)_Y$ gauge bosons and fourteen
pseudo-Goldstone bosons at low energies. As originally formulated,
this theory contains quadratically divergent operators whose
na\"{\i}ve signs predict a saddle point in the radiative potential.
If this theory is regulated using little technicolor, the calculable radiative potential reproduces the na\"{\i}ve expectation. For this reason it is interesting to ask whether a UV
extension into non-square theory space might remedy this radiative
instability, especially since the qualitative structure of the UV extended radiative potential has no relation to the original na\"ive potential.  However, as we will see, non-square UV extensions of
the $SO(9)/(SO(4)\times SO(5))$ little Higgs model not only fail to
remove the saddle point, but exacerbate the situation by introducing
a pseudo-Goldstone tadpole. The tadpole points towards
a stable vacuum that preserves all of $F$ and is thus inconsistent
with electroweak symmetry breaking.

The presence of a tadpole can be understood more broadly as a
generic property of non-square UV extensions.  These extensions generically suffer from
tadpole instabilities unless the link field vevs and the quadratic
Casimir of $F$ satisfy a particular relationship.  Consider two three-site UV extensions of
a NL$\Sigma$M with the global symmetry breaking pattern of
Eq.~(\ref{eq:masterbreak}).  The first is defined analogously to Eq.~(\ref{eq:Areg}) by
\begin{equation} \mbox{Theory A$_{\rm reg}$: } \qquad
\begin{tabular}{c}\includegraphics[clip,scale=0.4]{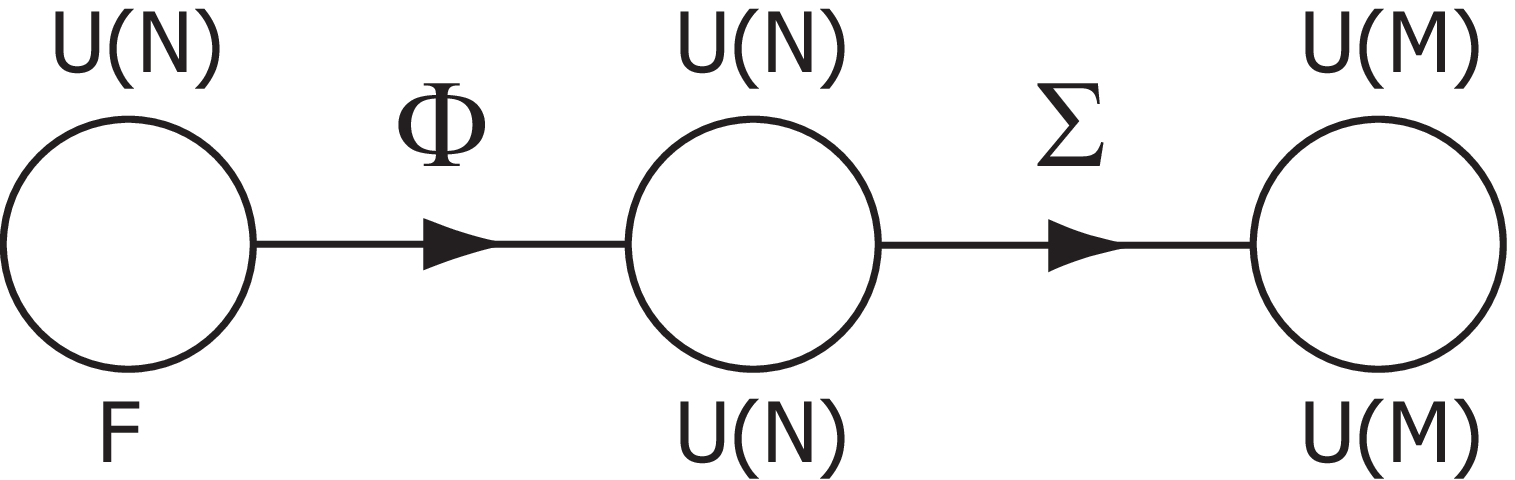}\end{tabular}
\end{equation}
\begin{eqnarray} \langle\Phi\rangle=\left(\begin{array}{c} \mathbf{1}_{N\times
N}\end{array}\right), & \qquad & \langle\Sigma\rangle =
\left(\begin{array}{c}
\mathbf{1}_{M\times M} \\
\mathbf{0}_{(N-M)\times M} \end{array}\right),
\label{eq:FNM_Moose1}\end{eqnarray} and the second analogously to Eq.~(\ref{eq:Breg})by
\begin{equation} \mbox{Theory B$_{\rm reg}$: } \qquad
\begin{tabular}{c}\includegraphics[clip,scale=0.4]{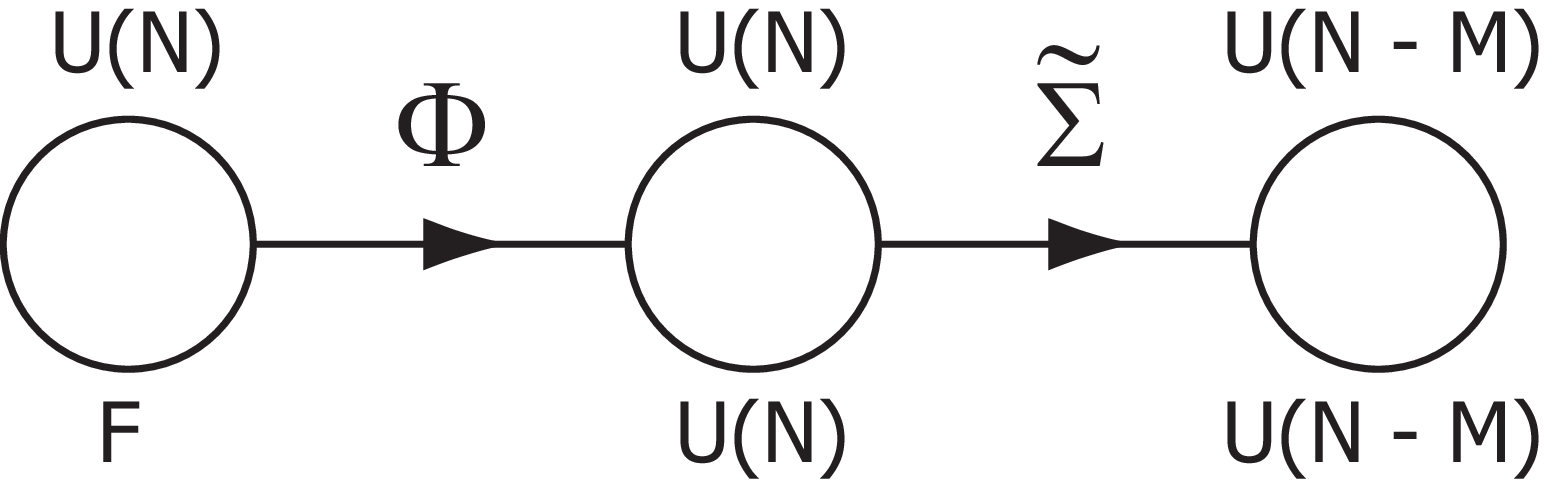}\end{tabular}
\end{equation}
\begin{eqnarray} \langle\Phi\rangle=\left(\begin{array}{c} \mathbf{1}_{N\times
N}\end{array}\right), & \qquad & \langle\tilde{\Sigma}\rangle =
\left(\begin{array}{c}
\mathbf{0}_{M\times (N-M)} \\
\mathbf{1}_{(N-M)\times (N-M)} \end{array}\right).
\label{eq:FNM_Moose2}\end{eqnarray}

To ascertain whether a theory has a tadpole instability, one must
compute the radiative potential. However, it is more elucidating to
first consider the space of gauge invariant operators that can be
generated at leading order in the gauge couplings.
Since ${\rm Tr}(M^2)$ and ${\rm Tr}(M^4)$ are constant in both
Theory A$_{\rm reg}$ and Theory B$_{\rm reg}$, neither theory
receives nontrivial radiative corrections at $\mathcal{O}(g^2)$ or
$\mathcal{O}(g^4)$. At order $\mathcal{O}(g^6)$, there is only one
allowed gauge invariant operator, namely
\begin{eqnarray}
{\rm Tr}(\Phi \Sigma \Sigma^\dagger \Phi^\dagger C_F)& = &
\frac{2}{f_{\rm eff}} {\rm Tr}(\pi T)+\mathcal{O}(\pi^2) + {\rm
constant}, \label{eq:gaugeinvop}
\end{eqnarray} where we are using the link field parametrization from
Eq.~(\ref{eq:goldstonepar}) and
\begin{eqnarray} T & = &
i[\langle\Sigma\rangle\langle\Sigma\rangle^\dagger,\langle\Phi\rangle^\dagger
C_F \langle\Phi\rangle].
\end{eqnarray}
Since $\pi\in G/(F\cup H)$, this leading operator is
tadpole-free if and only if the projection of $T$ onto $G/(F\cup H)$
is zero, \emph{i.e.} if
\begin{equation}
T |_{G/(F\cup H)} =0. \label{eq:criterion}
\end{equation}
The only effect of toggling is to flip the sign of $T$, so toggling has no effect on whether this criterion
is satisfied. While it is not obvious whether theories that satisfy
this constraint are tadpole-free to all orders in the radiative
potential, this determination can be made via an explicit evaluation
of the Coleman-Weinberg potential.

For example, consider the $SO(9)\times (SO(4)\times SO(5))$ little
Higgs theory, which can be UV extended into Theory A$_{\rm reg}$ or
Theory B$_{\rm reg}$ if we replace unitary groups with orthogonal
groups and take $N=9$ and $M=4$:
\begin{equation}
\begin{array}{ccc}
G & \; = \; &  SO(9), \\
H & \; = \; & SO(4)  \times SO(5), \\
F & \; = \; & SU(2)^3 \times U(1).
\end{array}
\end{equation}
To evaluate Eq.~(\ref{eq:CW2}) and Eq.~(\ref{eq:criterion}), we use
the Goldstone and gauge group embedding described in
\cite{LittlestHiggsModelwithCustodialSU(2)Symmetry}, rotated to a
basis where the link field vev is diagonal.  The fourteen
pseudo-Goldstone bosons in this theory comprise a Higgs doublet
$\vec{h}$, a singlet $\psi^0$, and three triplets $\psi^{ab}$, where
$a,b=1,2,3$. We represent $\vec{h}$ as a vector of $SO(4)\simeq
SU(2)_L \times SU(2)_R$ where $SU(2)_R$ is a custodial symmetry, and
group $\psi^0$ and $\psi^{ab}$ into a four by four matrix
\begin{equation} \Psi = \psi^0 \mathbf{1}_{4\times 4} + 8 \psi^{ab}T_L^a T_R^b,
\end{equation}
where $T_L^a$ and $T_R^b$ are generators of $SU(2)_L$ and $SU(2)_R$.
Going to unitary gauge, Theories A$_{\rm reg}$ and B$_{\rm reg}$
have link fields defined by Eqs.~(\ref{eq:goldstonepar}) and
(\ref{eq:feff}), where
\begin{eqnarray} \pi & = & -\frac{i}{4}\left(\begin{array}{cc}
\mathbf{0}_{4\times 4} & \begin{array}{cc} \Psi & 2 \vec{h} \end{array} \\
\begin{array}{c} -\Psi \\ -2 \vec{h}^T \end{array} & \mathbf{0}_{5\times 5}
\end{array}\right).
\end{eqnarray}
Also,
\begin{eqnarray} \langle\Sigma\rangle \langle\Sigma\rangle^\dagger
= \left(\begin{array}{cc} \mathbf{1}_{4\times 4} &
\mathbf{0}_{4\times 5} \\
\mathbf{0}_{5\times 4} & \mathbf{0}_{5\times 5}
\end{array}\right),\qquad C_F=
\frac{1}{8}\left(\begin{array}{cc}
\begin{array}{cc} \mathbf{5}_{4\times 4} & -\mathbf{1}_{4\times 4} \\
-\mathbf{1}_{4\times 4} & \mathbf{5}_{4\times 4}
\end{array} & \mathbf{0}_{8\times 1} \\
\mathbf{0}_{1\times 8} & 0
\end{array}\right).
\end{eqnarray}
Plugging these expressions into Eq.~(\ref{eq:criterion}), $T$ has a component in the direction of $\psi^0$, so the
singlet has a tadpole. We also verified this result by evaluating
the Coleman-Weinberg potential numerically.

A tadpole indicates that we are expanding around the wrong vacuum.
Thus it is natural to ask, does the tadpole point towards a
radiatively stable vacuum that is phenomenologically
viable?\footnote{Since the singlet transforms trivially under
electroweak gauge transformations, we know that \emph{any} vacuum
rotated into the direction of the singlet preserves (at least) the
$SU(2)_L \times U(1)_Y$ of the standard model. The only question is
whether such a vacuum preserves \emph{more} than this gauge
symmetry.} Denoting the singlet generator by $\Delta \in
G/(F\cup H)$, then rotating the vevs in this direction yields
\begin{equation}
\langle\Phi\rangle\rightarrow e^{i \Delta}\langle\Phi\rangle, \qquad
\langle\Sigma\rangle\rightarrow e^{i \Delta}\langle\Sigma\rangle,
\qquad \langle\tilde{\Sigma}\rangle \rightarrow e^{i
\Delta}\langle\tilde{\Sigma}\rangle.
\end{equation}
Computing Eq.~(\ref{eq:gaugeinvop}) for Theories A$_{\rm reg}$ and
B$_{\rm reg}$ to quadratic order in the singlet,
\begin{eqnarray}
{\rm Tr}(\Phi \Sigma \Sigma^\dagger \Phi^\dagger C_F) & = &
-\frac{\sin(\Delta)}{4}\left(\frac{\psi^0}{ f_{\rm
eff}}-\cot(\Delta)\right)^2 + {\rm constant},
\label{eq:custpot}
\end{eqnarray}
for Theory A$_{\rm reg}$ and likewise for Theory B$_{\rm reg}$
except with an overall minus sign. Thus, the leading operator is tadpole-free
if and only if
\begin{equation}
\Delta= \pi \left(n+\frac{1}{2}\right),\qquad n\in\mathbb{Z}
\end{equation}
From Eq.~(\ref{eq:custpot}) we see that $\psi^0$ has a potential
consisting of alternating local minima and maxima, each of which
leaves the entirety of $F$ unbroken. Toggling between
Theories A$_{\rm reg}$ and B$_{\rm reg}$ does little more than
interchange minima and maxima, so neither theory has the appropriate
low energy gauge group for successful electroweak symmetry breaking.
Again, we verified these results beyond the leading order by
computing the radiative potential numerically.  Thus, we conclude
that despite the control afforded by toggling, non-square UV
extensions of the $SO(9)\times (SO(4)\times SO(5))$ little Higgs
theory lack radiatively stable vacua consistent with
electroweak physics.
\section{Future Directions}
\label{sec:conclusion}

Understanding vacuum alignment is crucial for constructing realistic
theories based on spontaneous symmetry breaking.  While there is a
large variety of symmetry breaking patterns available for model
building, only certain vacuum alignments are radiatively stable in
the presence of gauge interactions.  We have shown that for a
certain class of phenomenologically interesting NL$\Sigma$Ms,
different ultraviolet physics can yield different stable vacuum
alignments in the infrared.  Unfortunately, our technique fails to
rectify vacuum instabilities in the $SO(9)/(SO(4)\times SO(5))$
 little Higgs model \cite{LittlestHiggsModelwithCustodialSU(2)Symmetry}.


Our result relies on using non-square mooses to UV extend
NL$\Sigma$Ms.  Like the technique of hidden local symmetry,
non-square UV extensions can have calculable radiative corrections,
but because they lack an extra-dimensional interpretation,
non-square mooses suggest new avenues for regulating more general
NL$\Sigma$Ms. As an interesting counterexample to the common lore,
the stable vacuum does not necessarily preserve
the maximal unbroken gauge symmetry in non-square mooses.

To what extent is it possible to engineer vacuum alignments in more
general effective field theories, including those which lack a
non-square moose representation?    In Section
\ref{sec:vendiagrams}, we argued that any NL$\Sigma$M can be UV
extended by adding (or subtracting) either unbroken global
symmetries or fully gauged broken symmetries. Are there other
realizations of this scenario? One might conjecture that non-square constructions could be generalized  to any $G/H$ NL$\Sigma$Ms where $H$ is a
maximal subgroup of $G$.\footnote{Indeed, the reason why we used $U(N)$ instead of $SU(N)$ groups is that a maximal
subgroup of $SU(N)$ is $SU(N-M)\times SU(M) \times U(1)$. The extra
$U(1)$ factor complicates toggling, though similar
results hold.}  More phenomenologically relevant would be a method to UV extend the $SU(6)/Sp(6)$  little Higgs \cite{LHsfromanAntisymmetricCondensate} to see whether the correct vacuum alignment for electroweak symmetry breaking could be ensured.


Another interesting question is how to UV complete non-square
link fields. The theory space link fields become strongly coupled at
$ \Lambda \sim 4\pi f$ so new physics is needed to restore unitarity
at that scale. Link fields can always be UV completed into linear
sigma models, but can they arise from strong dynamics? For example,
consider UV completing a non-square link field with a fermion
condensate $\langle \psi_{i} \psi_{j}^c \rangle$, where the $\psi_i$
($\psi_j^c$) transforms as a fundamental (anti-fundamental) under a
confining group $G_{S}$.  However, because the link field is
non-square by assumption, the number of fundamental and
anti-fundamental representations of $G_{S}$ are different,
generically introducing a gauge anomaly. One way of side-stepping
this anomaly is to include spectator fermions also charged under
$G_{S}$, a method considered for the $SU(4)/SU(3)$ sigma models in
the simple group little Higgs model \cite{LHfromaSimpleGroup}.
Alternatively, for non-square mooses with $SO(N)$ flavor symmetries,
one might look for a $G_{S}$ that exhibits confinement from two
different real representations.

Finally, in this paper we have considered vacuum alignment in the
presence of gauge fields alone.  In any realistic composite Higgs
theory, there will be additional contributions from fermion loops,
and because of the large top Yukawa coupling, fermion loops can give
the dominant contribution to the radiative potential.  Because
toggling changes the gauge structure of the theory it will also
change the allowed fermion representations, so the action of
toggling on the fermion sector is not well-defined.  Then again, for
little Higgs theories in particular, the challenge to making fully
realistic theories is less the sign of fermion radiative corrections
as the magnitude of those corrections, and generically, one needs
some level of fine-tuning to get the correct electroweak scale
\cite{Casas:2005ev}.   Still, the fact that non-square mooses have
such counter-intuitive properties inspires us to search for other
novel mechanisms to adjust not only the sign but perhaps the
magnitude of NL$\Sigma$M radiative potentials.

\acknowledgments{We thank Nima Arkani-Hamed for inspiring this
project and Howard Georgi, Can Kilic, and Aaron Pierce for
constructive comments on the manuscript.  This work is supported by
the DOE under contract DE-FG02-91ER40654.}

\appendix

\section{Two-Site Non-Square Mooses}
\label{sec:appendix1}

In this appendix, we show that the little
technicolor construction and Theories A and B from Section
\ref{sec:twosite} have the same light degrees of freedom as the
original NL$\Sigma$M.

The little technicolor theory is
defined in Eq.~(\ref{eq:LTC}). The global symmetry breaking pattern
is $U(N)_{L}\times U(N)_{R}\rightarrow U(N)_{V}$, which generates a
$U(N)$'s worth of Goldstone bosons. Since the Goldstone modes
parameterize the broken global symmetry directions, we define
$\Sigma=L\langle\Sigma\rangle R^{\dagger}$, where $(L,R)\in
U(N)_{L}\times U(N)_{R}$. The link field can be written as
\begin{equation} \Sigma=U\langle\Sigma\rangle,\end{equation} for some unitary matrix
$U$, which can in turn be expressed as the product of an element of
$H = U(M)\times U(N-M)$ and an element of $U(N)/H$
\begin{eqnarray} U
 =  e^{i\pi/f}e^{ih/f},\qquad \pi  =
\left(\begin{array}{cc}
0 & b\\
b^{\dagger} & 0\end{array}\right),\qquad h  =
\left(\begin{array}{cc}
a & 0\\
0 & c\end{array}\right),
\label{eq:cosetdecomp}
\end{eqnarray}
 where $a$ and $c$ are Hermitian $M\times M$ and $(N-M)\times(N-M)$
Goldstone matrices and $b$ is a general complex $M\times(N-M)$
Goldstone matrix. Since the $\Sigma$ vev breaks all of $H$, we can go to a
unitary gauge where the $h$ Goldstones are eaten, leaving
\begin{equation}
U=e^{i\pi/f}.
\label{eq:Udef}
\end{equation}
A straightforward calculation shows that the Lagrangian for the
little technicolor theory is
\begin{equation}
\mathcal{L}  =  -\frac{1}{2g_{F}^{2}}\Tr(F_{\mu\nu}F^{\mu\nu})-\frac{1}{2g_{H}^{2}}\Tr(H_{\mu\nu}H^{\mu\nu})+f^{2}\Tr|G_{\mu}+H_{\mu}|^{2},
\end{equation}
\begin{equation}
G_{\mu}  =  iU^{\dagger}D_{\mu}U, \qquad D_{\mu}  =  \partial_{\mu}+i F_{\mu},
\end{equation}
where $F_{\mu}$
and $F_{\mu\nu}$ denote the gauge field and field strength tensor
for the gauge group $F$, and analogously for $H$.  For later convenience
we define\begin{eqnarray} G_{\mu} & = & \left(\begin{array}{cc}
A_{\mu} & B_{\mu}\\
B_{\mu}^{\dagger} & C_{\mu}\end{array}\right).\end{eqnarray}


If we take the gauge coupling $g_{H}\rightarrow\infty$, then the
$H_{\mu}$ gauge bosons become ultra-massive and can be integrated
out by setting them to their equations of motion. Since $H_{\mu}$
has no kinetic term in this limit, it acts as a Lagrange
multiplier which effectively eliminates $A_{\mu}$ and $C_{\mu}$,
leaving the effective Lagrangian\begin{eqnarray} \mathcal{L} & = &
-\frac{1}{2g_{F}^{2}}\Tr(F_{\mu\nu}F^{\mu\nu})+f^{2}{\rm
Tr}\left|\left(\begin{array}{cc}
0 & B_{\mu}\\
B_{\mu}^{\dagger} & 0\end{array}\right)\right|^{2},\\
 & = & -\frac{1}{2g_{F}^{2}}\Tr(F_{\mu\nu}F^{\mu\nu})+f^{2}\Tr(p^{\mu}p_{\mu}) \label{eq:CCWZlang},\end{eqnarray}
where $p_{\mu}$ is the component of $G_\mu$ that falls
in the $U(N)/H$ direction.  Recall that Eq.~(\ref{eq:CCWZlang}) is
simply the CCWZ Lagrangian
\cite{StructureofPhenomenologicalLagrangiansI,StructureofPhenomenologicalLagrangiansII}
for a $U(N)/H$ NL$\Sigma$M with $F \subset U(N)$ gauged.  We have
arrived at the known result that little technicolor
reproduces the original NL$\Sigma$M at low energies.  For
finite $g_H$, the Lagrangian is the same as Eq.~(\ref{eq:CCWZlang})
with higher dimension operators suppressed by the mass of
the heavy $H$ gauge bosons \cite{LittleTC}.

Next, consider Theory A defined in Eq.~(\ref{eq:TI}). The
global symmetry breaking pattern is $U(N)_{L}\times
U(M)_{R}\rightarrow U(M)_{V}\times U(N-M)_{L}$, yielding a
$U(N)/U(N-M)$'s worth of Goldstones. Like before, we define
$\Sigma=L\langle\Sigma\rangle R^{\dagger}$, although this time
$(L,R)\in U(N)_{L}\times U(M)_{R}$. Since $\langle\Sigma\rangle$ is
a $N$ by $M$ matrix where $N>M$, we can ``pull'' $R^{\dagger}$
through the vev and write
$\Sigma=L\overline{R}^{\dagger}\langle\Sigma\rangle$, where
$\overline{R}^{\dagger}$ is defined by \begin{eqnarray} \overline{R}
& = & \left(\begin{array}{cc} R & \begin{array}{c}
0\end{array}\\
0 & 1\end{array}\right).\end{eqnarray}
 In general, if $\Sigma$ has more rows than columns, it is always
possible to write it as
\begin{equation}
\Sigma = U \langle\Sigma\rangle,
\label{eq:linkdef}
\end{equation}
where $U=L\overline{R}^\dagger$. If $\Sigma$ has fewer rows than
columns, then this parametrization can instead be made for
$\Sigma^{\dagger}$.

We perform the same coset decomposition of $\Sigma$ as in Eq.~(\ref{eq:cosetdecomp}). In this parametrization $c$ is eliminated
immediately by the vev and $a$ is eaten by the $H = U(M)$ gauge
bosons, leaving precisely Eq.~(\ref{eq:Udef}).  Since the
symmetries corresponding to $c$ are left unbroken by the vacuum,
there are no propagating $c$ Goldstone bosons.  Thus, we have shown
that Theory A has precisely the same Goldstone content as little
technicolor and the original NL$\Sigma$M, albeit through a slightly
different mechanism.

The Lagrangian for Theory A is\begin{eqnarray} \mathcal{L} & = &
-\frac{1}{2g_{F}^{2}}{\rm
Tr}(F_{\mu\nu}F^{\mu\nu})-\frac{1}{2g_{H}^{2}}{\rm
Tr}(H_{\mu\nu}H^{\mu\nu})+f^{2}{\rm
Tr}|G_{\mu}+\overline{H}_{\mu}|^{2}P_{\Sigma},\end{eqnarray} where
$G_{\mu}$ is defined as before,
$P_{\Sigma}=\langle\Sigma\rangle\langle\Sigma\rangle^{\dagger}$, and
$\overline{H}_{\mu}$ is an $N\times N$ matrix with $H_{\mu}$ in the
upper $M\times M$ block and zeroes in the lower $(N-M)\times(N-M)$
block. By sending $g_{H}\rightarrow\infty$ and integrating out
$H_{\mu}$, we eliminate just the $A_{\mu}$ component of $G_{\mu}$.
Thus, the effective Lagrangian is\begin{eqnarray} \mathcal{L} & = &
-\frac{1}{2g_{F}^{2}}\Tr(F_{\mu\nu}F^{\mu\nu})+f^{2}{\rm
Tr}\left|\left(\begin{array}{cc}
0 & B_{\mu}\\
B_{\mu}^{\dagger} & C_{\mu}\end{array}\right)\right|^{2}P_{\Sigma},\\
 & = & -\frac{1}{2g_{F}^{2}}\Tr(F_{\mu\nu}F^{\mu\nu})+\frac{1}{2}f^{2}\Tr(p^{\mu}p_{\mu}).
\label{eq:EffL}
 \end{eqnarray}
Note that $P_\Sigma$ eliminates the $C_\mu$ component from the
Lagrangian. Having again generated the CCWZ Lagrangian, we see that
little technicolor and Theory A have identical low energy physics up
to a factor of $\sqrt{2}$ in the pion decay constant.  The
higher dimensional operators generated from integrating out $H_\mu$
with finite $g_H$ are generically different from little technicolor.

A completely analogous computation can be done for Theory B, defined
in Eq.~(\ref{eq:TII}). We can express $\Sigma$ in the form of
Eq.~(\ref{eq:cosetdecomp}), although this time $c$ is
eliminated by the vev and $a$ is eaten.  After going to unitary
gauge the link field can be written as
\begin{equation}
\tilde{\Sigma}=U\langle\tilde{\Sigma}\rangle
\end{equation}
where $U$ is the same as in Eq.~(\ref{eq:Udef}).  Thus, we conclude
that the uneaten Goldstones in Theories A and B are exactly the
same. Moreover, by integrating out massive gauge bosons we find that
the effective Lagrangian for Theory B is again Eq.~(\ref{eq:EffL}).

\section{Parameters in the Simple Example}
\label{sec:appendix2}



Here, we define the parameters in the radiative
potentials for Theories A$_{\rm reg}$ and B$_{\rm reg}$ from
Section \ref{sec:threesite}.    To leading order in $g_F$, the gauge boson masses in Theory A$_{\rm
reg}$ are
\begin{eqnarray}
m_1^2 & = & s + \sqrt{s^2 - t^2}  + \mathcal{O}(g_F^2),\\
m_2^2 & = &  \frac{1}{2} f_\Phi^2 g_G^2 + \frac{1}{4} f_\Sigma^2
g_G^2,\\
m_3^2 & = &  s - \sqrt{s^2 - t^2} + \mathcal{O}(g_F^2),
\end{eqnarray}
where
\begin{equation}
s = \frac{1}{4} f_\Phi^2 g_G^2 +\frac{1}{4} f_\Sigma^2 g_G^2 +
\frac{1}{4}f_\Sigma^2 g_H^2, \qquad t = \frac{1}{2} f_\Phi f_\Sigma
g_G g_H.
\end{equation}  Note that $m_1^2>m_2^2>m_3^2$ for small enough $g_F$.  The coefficients $\alpha_i$ are defined as
\begin{equation}
\alpha_1 = \frac{1}{4}\frac{m_1^2 f_\Phi^2 g_G^2 + 2 t^2}{(m_1^2 -
m_2^2)(m_1^2 - m_3^2)},
\end{equation}
and similarly for $\alpha_2$ and $\alpha_3$.  For Theory B$_{\rm reg}$, we define
\begin{eqnarray}
\tilde{m}_1^2 & =&  \frac{1}{2} f_\Phi^2 g_G^2 + \frac{1}{4} f_\Sigma^2 g_G^2,\\
\tilde{m}_2^2  &=& \frac{1}{2}f_\Phi^2 g_G^2 + \mathcal{O}(g_F^2),
\end{eqnarray}
and for small enough $g_F$, $\tilde{m}_1^2 > \tilde{m}_2^2$.

\section{Vacuum Alignment in Doubly Non-square Mooses}
\label{sec:appendix3}

For NL$\Sigma$Ms of the form
\begin{equation}
\begin{array}{ccccc}
G & \; = \; & U(N)_{L}& \times &U(N)_{R},\\
H & \; = \; & U(P)_{L}\times U(N-P)_{L}&\times &U(M)_{R}\times
U(N-M)_{R},\\
F & \; = \; & &U(N)_{V},
\end{array}
\label{eq:NMPsymm}
\end{equation}
where $N>M,P$, it is possible to understand toggling and vacuum alignment completely analytically.\footnote{If $M$ or $P$ is greater than $N$, then all
Goldstone bosons are eaten and there is no meaning to vacuum
alignment.}    Such a theory is a natural
generalization of the simple group little Higgs
\cite{LHfromaSimpleGroup}, which apart from extra $U(1)$ factors is defined by
$N=3$, and $M=P=1$.

It is possible to UV extend the NL$\Sigma$M in
Eq.~(\ref{eq:NMPsymm}) into a UV insensitive theory by using two non-square link fields:
\begin{equation} \mbox{Theory A: } \qquad
\begin{tabular}{c}\includegraphics[clip,scale=0.4]{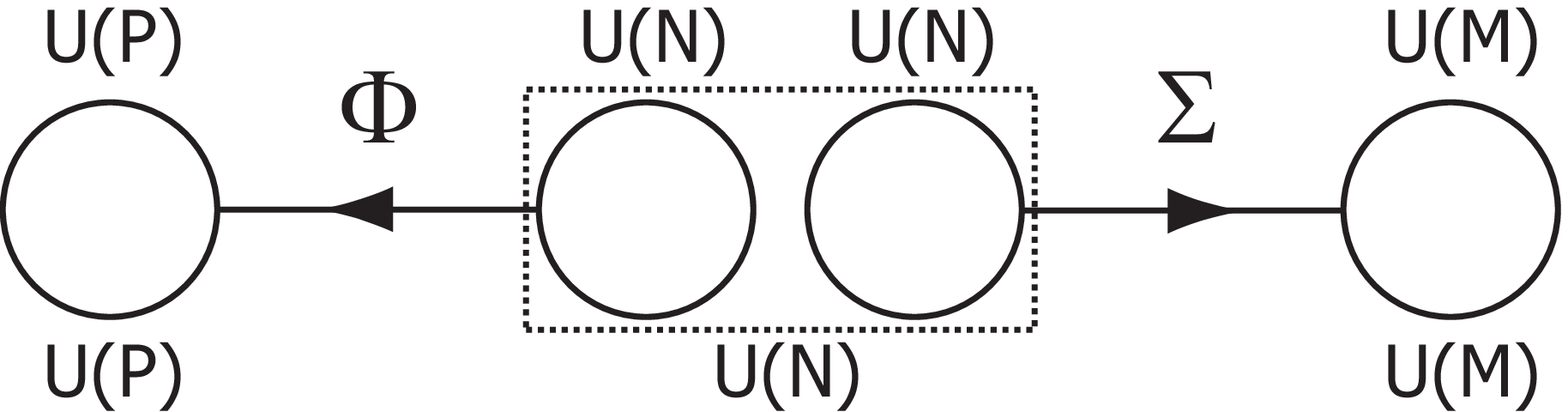}\end{tabular}\label{eq:genmoose1}
\end{equation}
\begin{eqnarray}
\langle\Sigma\rangle & = & \begin{array}{c} \left(\begin{array}{c}
\begin{array}{c}
\mathbf{1}_{M\times M}\\
\mathbf{0}_{(N-M)\times M}\end{array}\end{array}\right).\end{array}
\label{eq:genvev1}
\end{eqnarray}
The gauge couplings ($g_P$, $g_N$,
and $g_M$) and the pion decay constants ($f_\Sigma$ and $f_\Phi$)
can take arbitrary values.  We can toggle Theory A to Theory B by sending $U(M)
\rightarrow U(N-M)$
\begin{equation} \mbox{Theory B: } \qquad
\begin{tabular}{c}\includegraphics[clip,scale=0.4]{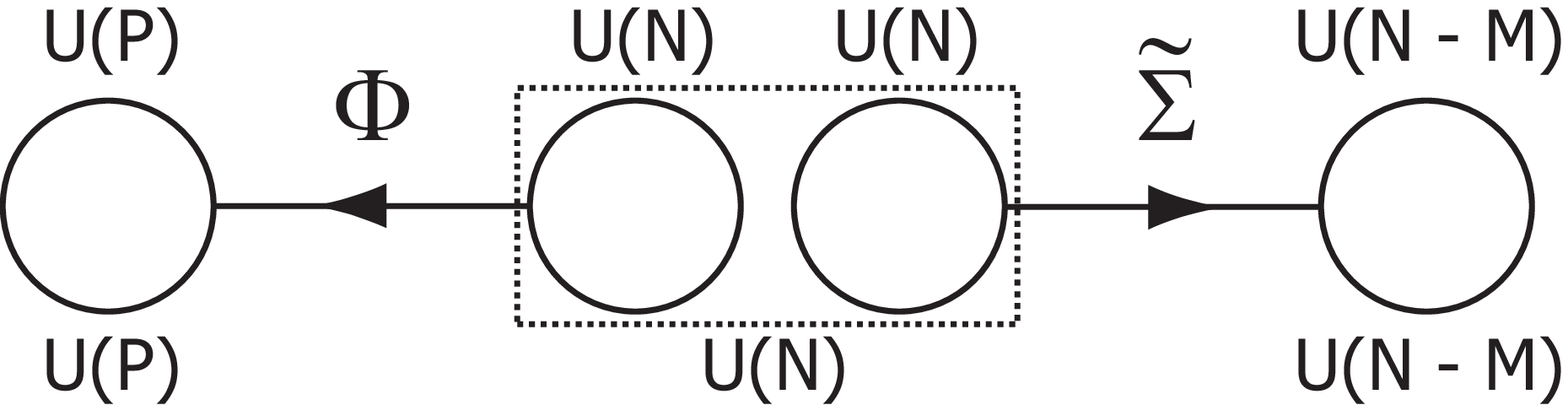}\end{tabular}
\end{equation}
\begin{eqnarray}
\langle\tilde{\Sigma}\rangle & = & \left(\begin{array}{c}
\begin{array}{c}
\mathbf{0}_{M\times(N-M)}\\
\mathbf{1}_{(N-M)\times(N-M)}\end{array}\end{array}\right),
\label{eq:genvev2}
\end{eqnarray}
yielding a second UV extension.  By toggling the $U(P)$
site it is possible to generate two additional UV extensions, Theories
C and D.  It is straightforward to see that all four theories have the same low energy degrees of freedom; they are comprised of a pair of two-site mooses with the
diagonal $U(N)$ weakly gauged, so we can simply invoke the low
energy equivalence between two-site mooses established in Section
\ref{sec:twosite}.

The stable vacuum alignment for each of these theories
can be determined from the radiative potential from
Eq.~(\ref{eq:CW}). By simply writing down the
leading gauge invariant operators, we see immediately that the leading order contribution comes at
$\mathcal{O}(g^4)$ and is
\begin{equation}
\Tr|\Phi^{\dagger}\Sigma|^{2}.
\end{equation}
Evaluating Eq.~(\ref{eq:CW}) explicitly for Theory A
yields\footnote{Interestingly, the logarithmic divergence is
proportional to the $U(N)$ gauge coupling only.  In this way,
non-square mooses are a generalization of the simple group method
for achieving collective symmetry breaking
\cite{LHfromaSimpleGroup}.}
\begin{equation}
V_{\rm CW} = -\frac{3 N}{256\pi^2} f_\Phi^2 f_\Sigma^2 g_N^4 \log
\left(\frac{\Lambda^2}{m^2} \right) {\rm Tr}|\Phi^\dagger \Sigma|^2,
\end{equation}
where $m$ is the scale of masses for the heavy gauge fields.

What is the stable vacuum alignment for non-square
mooses? For convenience, we construct projection operators
from the link field vevs by
\begin{equation} P_{\Phi} =
\langle\Phi\rangle\langle\Phi\rangle^{\dagger}, \qquad P_{\Sigma} =
\langle\Sigma\rangle\langle\Sigma\rangle^{\dagger}.
\end{equation}
Using the Goldstone parametrization of
Eq.~(\ref{eq:goldstonepar}), we can go to a unitary gauge where
\begin{eqnarray} \pi  = \{ P_{\Phi},\pi\} = \{
P_{\Sigma},\pi\}.
\end{eqnarray}
Expanding $V_{\rm CW} \sim - f_\Phi^2 f_\Sigma^2
\Tr|\Phi^{\dagger}\Sigma|^{2}$ to quadratic order in the Goldstones:
\begin{eqnarray}
V_{\rm CW}  &\sim&  - f_\Phi^2 f_\Sigma^2 \Tr\left(P_{\Phi}e^{i\pi/f_{\rm eff}} P_{\Sigma} e^{-i\pi/f_{\rm eff}}\right),\\
 & \sim & (f_\Phi^2 f_\Sigma^2/f_{\rm eff}^2)\Tr\left(\pi^{2}(2P_{\Phi}P_{\Sigma}-P_{\Sigma})\right)+\mathcal{O}(\pi^{3}),\\
 & \sim & (f_\Phi^2 f_\Sigma^2/f_{\rm eff}^2)\Tr\left(\pi^{2}(2P_{\Phi}P_{\Sigma}-P_{\Phi})\right)+\mathcal{O}(\pi^{3}).
 \end{eqnarray}
 Thus, the theory is tachyon-free if and only if
 \begin{equation}
 \label{tachfreecrit}
 P_{\Phi}P_{\Sigma}=P_{\Sigma} \qquad
\mbox{or} \qquad P_{\Phi}P_{\Sigma}=P_{\Phi}.
\end{equation} Immediately we see that the more
``aligned'' the vevs are, the more likely that the projection
operators constructed from them will satisfy the tachyon-free
constraint.

\FIGURE[t]{\epsfig{file=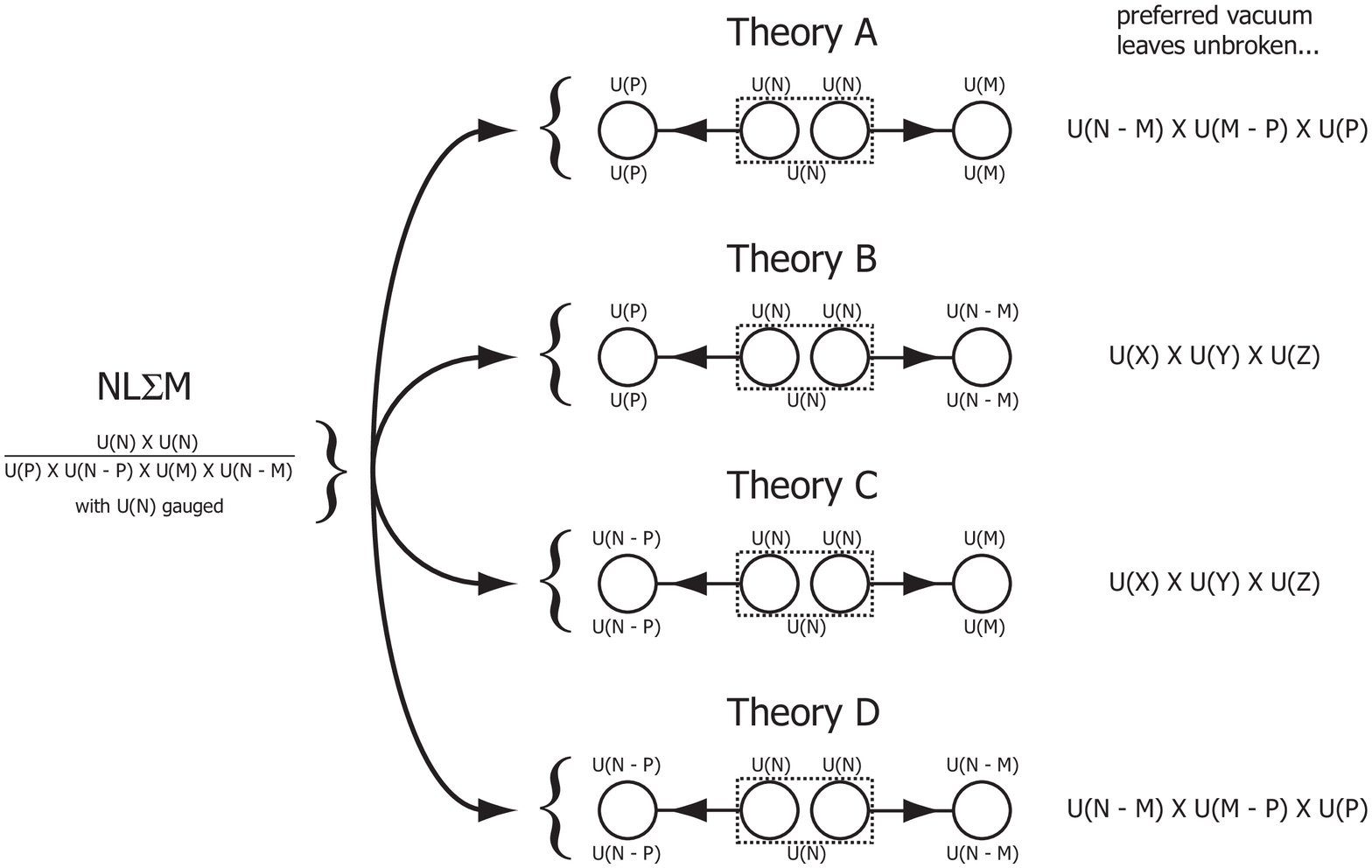,scale=0.5}
        \caption{Given the NL$\Sigma$M defined in Eq.~(\ref{eq:NMPsymm}), one can generate four different UV extensions based on non-square mooses.  The four theories yield two different stable vacuum alignments, and the stable vacuum is the one for which the link field vevs are maximally parallel.  Without loss of generality, we assume  $N> M \geq P$, and we define $X = \min(M,N-P)$, $Y = \min(P,N-M)$ and $Z= | N-M-P |$}
\label{MainPoint}}

We can use Eq.~(\ref{tachfreecrit}) to determine the stable vacuum
alignment in generic non-square mooses.  For the theory in
Eq.~(\ref{eq:genmoose1}) we can assume $N>M\geq P$ without loss of
generality. If the theory is to be tachyon-free, then $P_\Phi$ must
lie entirely within $P_\Sigma$. Consequently, the stable vacuum
alignment leaves
\begin{equation}
\mathcal{F}\cap\mathcal{H} = U(N-M)\times U(M-P) \times U(P)
\end{equation}
as the unbroken gauge symmetry.

Putting together all these results, there are four different ways to
UV extend the NL$\Sigma$M in Eq.~(\ref{eq:NMPsymm}) using non-square
mooses.   All four are related to Theory A in
Eq.~(\ref{eq:genmoose1}) by toggling either $\Phi$ or $\Sigma$ or
both. As summarized in Figure \ref{MainPoint}, these four mooses
yield two different stable vacuum alignments: one is the na\"{\i}ve
vacuum alignment ``predicted'' by
\cite{AlignmentoftheVacuuminTheoriesofTC} and the other is a novel
vacuum alignment from the perspective of the original NL$\Sigma$M.

\end{document}